\documentclass[lettersize,journal]{IEEEtran}
\usepackage{amsmath,amsfonts}
\usepackage{color}
\usepackage{algpseudocode} 
\usepackage{algorithm}
\usepackage{multirow}
\usepackage{booktabs}
\usepackage{array}
\usepackage[caption=false,font=normalsize,labelfont=sf,textfont=sf]{subfig}
\usepackage{textcomp}
\usepackage{stfloats}
\usepackage{url}
\usepackage{verbatim}
\usepackage{graphicx}
\usepackage{xcolor}
\usepackage{breqn}
\usepackage[T1]{fontenc}
\usepackage{hhline}
\usepackage{enumitem}
\usepackage{listings}
\usepackage{amssymb}
\usepackage{ulem}
\usepackage{float}

\newcommand\Tstrut{\rule{0pt}{2.5ex}} 

\definecolor{dkgreen}{rgb}{0,0.6,0}
\definecolor{gray}{rgb}{0.5,0.5,0.5}
\definecolor{mauve}{rgb}{0.58,0,0.82}

\lstdefinestyle{base}{
  language=Python,
  emptylines=1,
  breaklines=true,
  basicstyle={\scriptsize\ttfamily},
  moredelim=**[is][\color{red}]{@}{@},
  keywordstyle=\color{blue},
  commentstyle=\color{dkgreen},
  aboveskip=3mm,
  belowskip=3mm,
  frame=tb,
  breaklines=false,
}

\def\BibTeX{{\rm B\kern-.05em{\sc i\kern-.025em b}\kern-.08em
    T\kern-.1667em\lower.7ex\hbox{E}\kern-.125emX}}
\usepackage{balance}
\begin{document}
\title{ApproxTrain: Fast Simulation of Approximate Multipliers for DNN Training and Inference}


\author{\IEEEauthorblockN{Jing Gong\IEEEauthorrefmark{1},
Hassaan Saadat\IEEEauthorrefmark{1},
Hasindu Gamaarachchi\IEEEauthorrefmark{3}\IEEEauthorrefmark{1}, 
Haris Javaid\IEEEauthorrefmark{4},\\
Xiaobo Sharon Hu\IEEEauthorrefmark{2} and
Sri Parameswaran\IEEEauthorrefmark{1}\\
}
\IEEEauthorblockA{\IEEEauthorrefmark{1}School of Computer Science and Engineering,
UNSW Sydney, Kensington NSW 2052 Australia}\\
\IEEEauthorblockA{\IEEEauthorrefmark{3}Garvan Institute of Medical Research, Darlinghurst NSW 2010 Australia}
\IEEEauthorblockA{\IEEEauthorrefmark{4}AMD, Singapore}\\
\IEEEauthorblockA{\IEEEauthorrefmark{2}Department of Computer Science and Engineering
University of Notre Dame, Notre Dame, IN 46556 USA}


}

\algdef{SE}[VARIABLES]{Variables}{EndVariables}
   {\algorithmicvariables}
   {\algorithmicend\ \algorithmicvariables}
\algnewcommand{\algorithmicvariables}{\textbf{global variables}}

\maketitle

\begin{abstract}

Edge training of Deep Neural Networks (DNNs) is a desirable goal for continuous learning; however, it is hindered by the enormous computational power required by training. Hardware approximate multipliers have shown their effectiveness for gaining resource-efficiency in DNN inference accelerators; however, training with approximate multipliers is largely unexplored. To build resource-efficient accelerators with approximate multipliers supporting DNN training, a thorough evaluation of training convergence and accuracy for different DNN architectures and different approximate multipliers is needed. This paper presents ApproxTrain\footnote{https://github.com/AaronJing/ApproxTrain}, an open-source framework that allows fast evaluation of DNN training and inference using simulated approximate multipliers. ApproxTrain is as user-friendly as TensorFlow  (TF) and requires only a high-level description of a DNN architecture along with C/C++ functional models of the approximate multiplier. We improve the speed of the simulation at the multiplier level by using a novel LUT-based approximate floating-point (FP) multiplier simulator on GPU (AMSim). Additionally, a novel flow is presented to seamlessly convert C/C++ functional models of approximate FP multipliers into AMSim. ApproxTrain leverages CUDA and efficiently integrates \textcolor{black}{AMSim into the TensorFlow library, in order to} overcome the absence of native hardware approximate multiplier in commercial GPUs. We use ApproxTrain to evaluate the convergence and accuracy of DNN training with approximate multipliers for small and large datasets (including ImageNet) using LeNets and ResNets architectures. The evaluations demonstrate similar convergence behavior and negligible change in test accuracy compared to FP32 and bfloat16 multipliers. Compared to CPU-based approximate multiplier simulations in training and inference, the GPU-accelerated ApproxTrain is more than 2500x faster. Based on highly optimized closed-source cuDNN/cuBLAS libraries with native hardware multipliers, the original TensorFlow is, on average, only 8x faster than ApproxTrain.







\end{abstract}


\section{Introduction}

\IEEEPARstart{T}{he} \textit{training} phase in deep learning is significantly more computationally demanding than the \textit{inference} phase. Recent works have shown the importance of moving training to the edge to perform continuous learning \cite{aledhari_federated_2020}, though such deployments are scarce due to the high training cost. Thus, training has so far been largely relegated to high-performance computers and the cloud.
To support training at the edge, besides the accuracy of the learning model, energy/power and area efficiency are paramount. 

An efficient DNN system could be realized through two distinct yet complementary approaches: (1) by exploring and finding efficient \textcolor{black}{DNN hardware implementation schemes} for both training and inference; and (2) by exploring and finding the most suitable DNN architecture for the given problem. 
\textit{This paper presents ApproxTrain, a framework that allows fast evaluation of \textcolor{black}{DNN training when} using a variety of simulated approximate multipliers.} ApproxTrain facilitates thorough evaluations of training of many different \textcolor{black}{DNNs} with differing approximate multipliers, in a user-friendly manner with practically feasible runtimes to find a suitable \textcolor{black}{approximate multipliers that can be integrated into edge devices} for continuous learning. 

\textcolor{black}{Multipliers are one of the most compute-intensive hardware elements within a deep learning system~\cite{Jain2018compensatedDNN}. Thus, using approximate multipliers becomes promising scheme} for efficient implementation of DNN systems
\cite{Georgios2021}, and the efficacy of various approximate multiplier designs in DNN inference in terms of model accuracy and implementation efficiency has been demonstrated in recent years~\cite{vaverka_tfapprox_2020}. However, DNN training using approximate multipliers is largely unexplored.
A major obstacle in the exploration and evaluation of approximate multipliers in DNN training is the absence of
native hardware support for customized approximate floating-point (FP) multipliers on commercial processing units (CPUs, GPU, NPUs).
Hence, software simulation of approximate multiplications is needed for handling DNN training with approximate multipliers.

Thorough and effective evaluation of approximate multiplier
designs in DNN training has two practical requirements:
(1) run-time of training simulation \textcolor{black}{should be} sufficiently
fast and practically feasible; and (2) the ability to describe DNN architectures at a high level
so that various DNN architectures can be quickly constructed and
evaluated.
However, resorting to software simulation makes it challenging to meet the above requirements.
Firstly, since training is time-consuming, \textcolor{black}{inefficient simulation may further slow down training} (up to orders of magnitude), 
making its run-time practically infeasible. 

Secondly, the standard frameworks, such as TensorFlow and Pytorch, that allow a high-level description of DNN architectures and take advantage of the computational power of commercial GPUs/TPUs/NPUs, are not equipped to utilize approximate multipliers, since they invoke the native multipliers (built in the hardware) of these commercial platforms. 
Moreover, these deep learning frameworks, although open-source themselves, are based on closed-source cuDNN and cuBLAS libraries~\cite{cudnnclosed} at their backend. 
Hence, any modification in these high-level frameworks for incorporating fast approximate multiplier simulation while preserving their flexibility, requires specialized parallel programming skills, making it a daunting task for \textcolor{black}{a regular DNN developer} or approximate multiplier designer.

ApproxTrain \textcolor{black}{overcomes the above challenges through three key contributions.}
First, to improve the speed of ApproxTrain at the multiplier-level, we use a mantissa lookup table-based approach for functional simulation of approximate multipliers.
Our \textcolor{black}{mantissa-lookup} approach is based on the observation that only mantissa is approximated in most state-of-the-art approximate multipliers~\cite{saadat_realm_2020,saadat_minimally_2018}. We propose to compute the sign and exponent using the standard approach whereas compute the mantissa using the lookup table, thus, allowing smaller lookup tables, instead of constructing a lookup table for full bit-width of the operands. 
The alternate method of simulating approximate FP multipliers by bit-manipulation causes significant GPU instruction overhead and more importantly, results in variable performance for different multipliers. The proposed mantissa lookup table-based approach also makes the speed independent of the type of approximate multipliers.

Any training mechanism based on CPU can be painfully slower compared to the standard GPU implementations. Therefore, our second contribution, aiming to improve system-level performance, is the development of GPU-accelerated custom CUDA kernels as an alternative for non-modifiable closed-source cuDNN and cuBLAS libraries. Our GPU-accelerated custom CUDA kernels exploit various architectural features of the GPU ((such as fine-grained and coarse-grained parallelism, memory coalescing and on-chip shared memory). 
To reduce the number of kernel invocations (thus improving performance) and the necessary memory footprint during the weight gradient computation, we exploit the nature of the dilation step to efficiently integrate it into the image to column (IM2COL) kernel.

Third, to preserve the flexibility and ease-of-use of high-level framework (TensorFlow), we create custom approximate TF \textit{ops} (explained in section \ref{sec_background}) and seamlessly integrate them into TensorFlow.
These custom \textit{TF} ops support different types of DNN layers with approximate multiplication. The GPU-accelerated custom CUDA kernels (described above) are used to implement our custom approximate TF \textit{ops}.
ApproxTrain requires only standard TensorFlow-based implementation of DNN architectures, along with C/C++-based functional models of the approximate multipliers.

{\textcolor{black}{Overall}, ApproxTrain is a DNN framework that: (1) allows fast DNN training evaluations with approximate multipliers; (2) is similar and as user-friendly as the popular high-level frameworks such as TensorFlow; and (3) simplifies the daunting task of integrating approximate multiplier simulation in TensorFlow (or similar frameworks) to make it transparent to the DNN architecture/multiplier designer.} To this end, the paper makes the following key \textbf{contributions}.

\begin{itemize} 
\item A novel flow is presented to seamlessly transform C/C++ functional models of any approximate FP multipliers provided by designers into a lookup table (LUT)-based simulator (AMSim). This LUT-based AMSim requires negligible GPU memory compared to a whole LUT-based method that stores all multiplication results into LUT.

\item We present ApproxTrain, a TensorFlow-based framework, that allows fast DNN training evaluations with AMSim (simulated approximate FP multipliers). ApproxTrain is as user-friendly as the TensorFlow framework and requires only a high-level description of a DNN architecture along with C/C++ functional model of the approximate multiplier. ApproxTrain also supports inference using approximate multipliers.

\item Since the closed-source cuDNN and cuBLAS libraries cannot be modified to integrate functional models of approximate multipliers into ApproxTrain, we developed GPU-accelerated custom CUDA kernels.

\item To explore prospects of DNN training with approximate multipliers, we use ApproxTrain to evaluate DNN training with approximate multipliers in terms of training convergence and accuracy results. Our experiments with small and large datasets (including ImageNet) using popular neural network architectures demonstrate successful convergence with negligible change in test accuracy. 

\item The ApproxTrain framework, as well as the developed custom CUDA libraries, are released as an open-source repository \cite{AMDNN}.

\end{itemize}

 There has been an attempt for such a framework. \textit{However, that framework only supports DNN inference with 8-bit approximate integer multipliers~\cite{vaverka_tfapprox_2020}}. \textcolor{black}{Thus, to our best knowledge, there is no general framework that supports fast and user friendly DNN training as well as inference with approximate FP multipliers, i.e., in both forward and backpropagation phases of training. }


\subsection*{Paper Organization:} 
Section~\ref{subsec_motivation} presents the motivation of using approximate multipliers in training.
The relevant background and related work for this paper are discussed in Section~\ref{sec_background} and Section~\ref{sec_relatedwork}, respectively.
Section~\ref{sec_ams} describes the novel LUT-based approximate FP multiplier simulation, and Section~\ref{sec_approxtrain} describes the integration of the novel simulation into the presented framework ApproxTrain.  Experimental evaluation setup using ApproxTrain is elaborated in Section~\ref{sec_expsetup}.
The training convergence and accuracy results using approximate multipliers are presented in Section~\ref{sec_results_accuracy}, and the evaluation of the run-time performance of ApproxTrain is presented in Section~\ref{sec_results_performance}.
Finally, Section~\ref{sec_conc} concludes the paper.

\section{Motivation: The Promise of Approximate Multipliers in DNN Training}
\label{subsec_motivation}

The training phase of DNN requires large dynamic range to represent the intermediate data, such as gradients~\cite{kuchaiev_mixed-precision_2018}. Therefore, the resource-hungry floating-point (which offers large dynamic range) is the widely used format for DNN training.
Until a few years ago, the IEEE single-precision floating-point (FP32: having 8-bit exponent, 23-bit mantissa) format was used in training to achieve best possible accuracy results~\cite{micikevicius_mixed_2017}.
Recently, Brain Floating Point format (bfloat16) for multiplication was introduced, which has similar dynamic range (8-bit exponent), but lower precision (7-bit mantissa) compared to the FP32 ~\cite{wang_bfloat16_2019}. The bfloat16 format is currently being supported by Google, Nvidia and Intel in their latest TPUs, GPUs and NPUs \cite{noauthor_nvidia_nodate, alberto_villarreal_cueva_intel_2020, wang_bfloat16_2019}.

\begin{figure}[!t]
\begin {center}
\includegraphics[width=0.80\columnwidth,trim={0 13cm 25.2cm 0cm},clip=true]{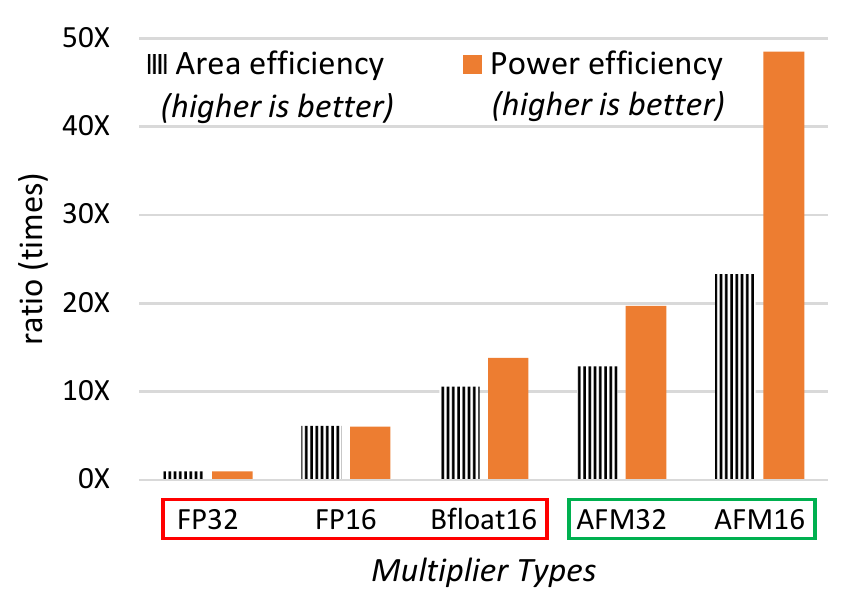}\\
\caption{Comparison of resource-efficiency (higher is better) of IEEE FP32, IEEE FP16, bfloat16 and approximate multipliers (AFM32 and AFM16). All area and power values are normalized with area and power of FP32, respectively. The multipliers are single cycle designs, and logic synthesis is done using Cadence RC compiler for TSMC 45nm cell library at 1GHz.}

 \label{fig2}
  \end {center}
\end{figure}

A comparison of resource-efficiency (higher is better) of IEEE FP32, FP16, and bfloat16 multipliers with the 32-bit and 16-bit versions of approximate FP multiplier from the literature (AFM32 and AFM16) \cite{saadat_minimally_2018} is shown in Figure~\ref{fig2}. It can be observed that the approximate multipliers are more power-efficient and more area-efficient than the FP32, FP16, and the bfloat16 multipliers.
Hence, we expect the approximate hardware multipliers to enable more resource-efficient training than when using FP32, FP16 or bfloat16 formats.  
However, before any custom neural accelerator (equipped with approximate hardware multipliers) can be built to exploit this promise, a thorough evaluation of training convergence and accuracy results for different DNN architectures is needed. This necessitates the fast evaluation framework for DNN training using approximate multipliers, which is the focus of this paper.

\section{background}
\label{sec_background}

\subsection{Deep Neural Networks}
\label{sec_bgdnn}

\begin{figure}[t]
    \centering
    \includegraphics[width=\linewidth]{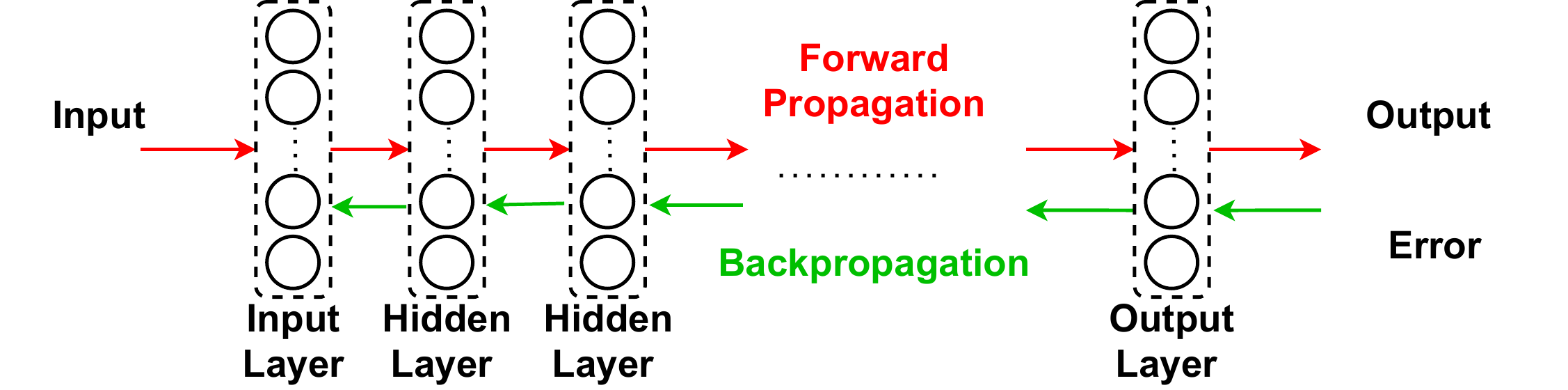}
    \caption{Forward propagation and backpropagation between DNN layers.}
    \label{fig:fpbplayer}
\end{figure}
\begin{figure}[t]
    \centering
    \includegraphics[width=\linewidth]{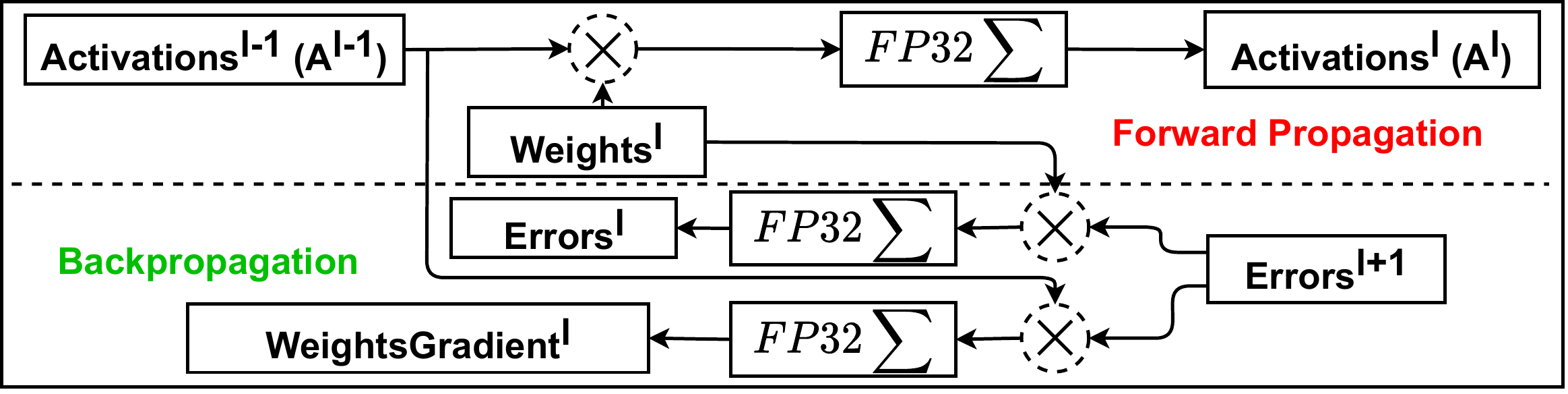}
    \caption{DNN training algorithm with two stages: forward propagation and backpropagation.}
    \label{fig:prop1}
\end{figure}
\begin{table}[t]
\centering
\setlength\tabcolsep{2.8pt}
\caption{Some popular neural network architectures and types of constituent layers \label{tab:commlayers}}
\begin{tabular}{|l|ccccc|}
\hline

DNN architecture & Dense & Convolution & \begin{tabular}[c]{@{}c@{}} \Tstrut Depthwise \\ Convolution\end{tabular} & Pooling & \begin{tabular}[c]{@{}c@{}} \Tstrut MultiHead\\ Attention\end{tabular} \\ 

\hline
\hline
\Tstrut LeNet-300-100       &    \checkmark   &      -       &           -                                                       &      -   &           -                                                    \\ \hline
\Tstrut ResNet      &     \checkmark   & \checkmark             &        -                                                          &  \checkmark        &    -                                                           \\ \hline
\Tstrut MobileNets  &    \checkmark    &     \checkmark         &           \checkmark                                                        &    \checkmark      &       -                                                        \\ \hline
\Tstrut Transformer &   \checkmark     &     -        &      -                                                            &  -       &    \checkmark                                                          \\ \hline
\end{tabular}
\end{table}

\textbf{DNN structure:} A DNN consists of an input layer, multiple hidden layers, and an output layer, as shown in Figure \ref{fig:fpbplayer}. Each layer is composed of multiple parallel neurons; for instance, the hidden layer in Figure \ref{fig:fpbplayer} has four parallel neurons. Each parallel neuron contributes to a weighted output that acts as an input for the next layer. Inputs and weights are subjected to linear algebra operations in each layer, followed by non-linear activation functions. Table \ref{tab:commlayers} shows common types of layers used in some common DNN architectures. Dense layer is used by all architectures listed since it can be used as a classification layer at the output of most DNN architectures. It can be simply used in MLPs (multi-layer perceptrons) as feedforward layers. The core of dense layer is matrix-vector multiplication. Convolution and depthwise convolution are widely used in image classification tasks, such as ResNet and MobileNets in Table \ref{tab:commlayers}. Convolution consists of multiplication-intensive operations.
Pooling layers are responsible for down-sampling, reducing the feature maps size, and not involving multiplications.
The MultiHeadAttention layer has shown extraordinary performance in language and image tasks, and the famous example is Transformer, as shown in Table \ref{tab:commlayers}. The MultiHeadAttention layer involves matrix multiplication under the hood.

\textbf{DNN Training:} {\color{black}Training is an iterative process that finds optimal parameters (weights) to reduce the difference between the model prediction and the dataset-labels. Initially, the weights are randomly generated based on a particular distribution. Then, a series of forward propagation and backpropagation phases are executed until no further accuracy improvement occurs. \textcolor{black}{Training is highly resource hungry and processes a large number of multiplications.}
}

The training procedure is shown as a flow diagram in Figure~\ref{fig:prop1}.
Activations from layer $l{-}1$ are multiplied with weights in the current layer $l$, and results are accumulated to become the activation for the next layer. After forward propagation, an error is calculated to reflect the difference between the predicted result and the label, known as propagation error. Propagation error is propagated backward to calculate gradients for weights in each layer so that the DNN model could achieve better prediction. As shown in Figure \ref{fig:prop1}, $Error^{l+1}$ is propagated backwards to layer $l$. To get $WeightsGradient^l$, multiply–accumulation is performed on $Errors^{l+1}$ and $Activations^{l-1}$. Then, $Weights^l$ could be updated with the calculated gradient. Similarly, $Errors^{l}$ (preceding layer gradient) is computed as the multiply-accumulation operation between the $Weights^{l}$ and the $Errors^{l+1}$ backpropagated from the succeeding layer~$l{+}1$.

\textbf{Training Accuracy \& Test Accuracy:}
The dataset for a neural network is divided into two subsets: a training set and a test set. The training set is used to train the neural network. Training accuracy is the classification accuracy calculated using the training set and is mostly used for monitoring training convergence. The test accuracy is one metric to evaluate the performance of the trained neural network when classifying the `unseen' test dataset.


{\color{black}
 \textbf{TensorFlow (TF):} TensorFlow, one of the most popular deep learning libraries, has been used to develop many DNNs across different application domains. TensorFlow provides highly abstracted \textit{Ops} such as Conv2D (2D Convolution, also known as Conv2D layer) and Dense (also known as fully connected layer) that are commonly used across different architectures and applications, allowing the users to build models easily. In TensorFlow, DNN architectures are represented as graphs, and the \textit{Ops} are nodes that take one or more tensors (multi-dimensional arrays) as inputs and perform computations on those tensors.
 Every \textit{Op} has its Compute method that defines the mathematical operation to be performed on the tensors. \textit{Ops} typically has a corresponding gradient \textit{Op} which is used in the backpropagation in the training phase. The backend of TensorFlow utilizes  \textit{cuBLAS} and \textit{cuDNN} libraries developed by NVIDIA for GPU acceleration. Both closed-source libraries are highly optimized low-level primitives for linear algebra and DNN. 
}
\subsection{Approximate Computing and Approximate Multipliers}

Many modern compute-intensive applications, including machine learning and DNN algorithms, are error-resilient: they can tolerate errors in underlying computations with negligible degradation in the final output quality~\cite{mittal_survey_2016}. Approximate computing aims to exploit this property by trading-off exactness in underlying computations for disproportionate gains in resource efficiency.

Hardware approximate multipliers (or approximate arithmetic units, in general) are resource-efficient computation units in which the hardware logic circuit is simplified, such
that they become faster, smaller, and/or less power/energy-hungry
while producing slightly erroneous outputs when compared to an exact multiplier.
Depending on the data type, approximate multipliers can be classified into an approximate integer or approximate FP multipliers. Approximate integer multipliers have been demonstrated to be effective in the inference phase of DNN. However, DNN training using approximate multipliers is largely unexplored. More details on approximate multipliers can be found in~\cite{saadat_minimally_2018}.

\section{Related Work}
\label{sec_relatedwork}

{Approximate hardware} for DNNs typically refers to using approximate arithmetic units for DNNs inference. In the computation of DNNs, multiplications dominate operations and consume the most power and area. Thus, utilizing approximate multipliers could improve inference efficiency and has been extensively studied \cite{saadat_minimally_2018,kim_efficient_2019,kim_low-cost_2021,shirane_design_2021}. For example, in~\cite{saadat_minimally_2018}, Saadat et al. replaced accurate multipliers with minimally biased multipliers in AlexNet during the inference stage. Other works \cite{kim_efficient_2019,kim_low-cost_2021,shirane_design_2021} all show significant energy savings when using approximate multipliers with minimal accuracy degradation. These works use only a limited variety of DNNs, support only inference, and do not report the run-time of simulations. TFapprox \cite{vaverka_tfapprox_2020} enables fast and flexible simulation of approximate multipliers in DNN inference due to TensorFlow integration and GPU acceleration. \textit{However, TFapprox is limited to 8-bit integer multiplications and supports inference only.}

Most of these efficiency gains were made for DNN inference. Few works focus on training DNNs with approximate multipliers. In~\cite{kim_efficient_2019}, Kim et al. claimed approximate multipliers are not suitable for training because DaDianNao~\cite{chen_dadiannao_2014} project failed to train the DNN with the fixed-point data type. Fixed-point 16-bit has a limited dynamic range that prevents some necessary small gradients from being represented~\cite{zhang_fixed-point_2020}. In \cite{hammad_deep_2019}, Hammad et al. attempted to train a DNN with approximate multipliers first and further improved convergence with accurate multipliers. \textit{However, only VGGnet was evaluated; thus, its feasibility is not shown on a wide variety of neural networks.} The work in \cite{cheng_logarithm-approximate_2020} first attempted to train neural networks with logarithm-based approximate multipliers. \textit{However, the evaluated model is simple: only a fully connected layer was considered}. A fully connected layer is well known to be redundant; for example, in~\cite{han_deep_2016}, Han et al. achieved a compression rate of 40 times for LeNet-300-100 (MLP), with parameters reducing from 1070kb to 27kb. Therefore, the convergence could have been well caused by redundancy. Given these concerns, it is challenging to show the efficacy of the work in~\cite{cheng_logarithm-approximate_2020}.

To overcome the above limitations, we first present a novel LUT-based approximate FP multiplier simulator on GPU (AMSim) that could efficiently simulate any type of approximate FP multipliers. Then, we integrate the AMSim into the ApproxTrain framework, supporting training and inference. Several neural network architectures with convolution and dense layers were evaluated, including training and testing on MNIST, CIFAR10, and ImageNet datasets.

\begin{figure}[t]
    \centering
    \includegraphics[width=\linewidth]{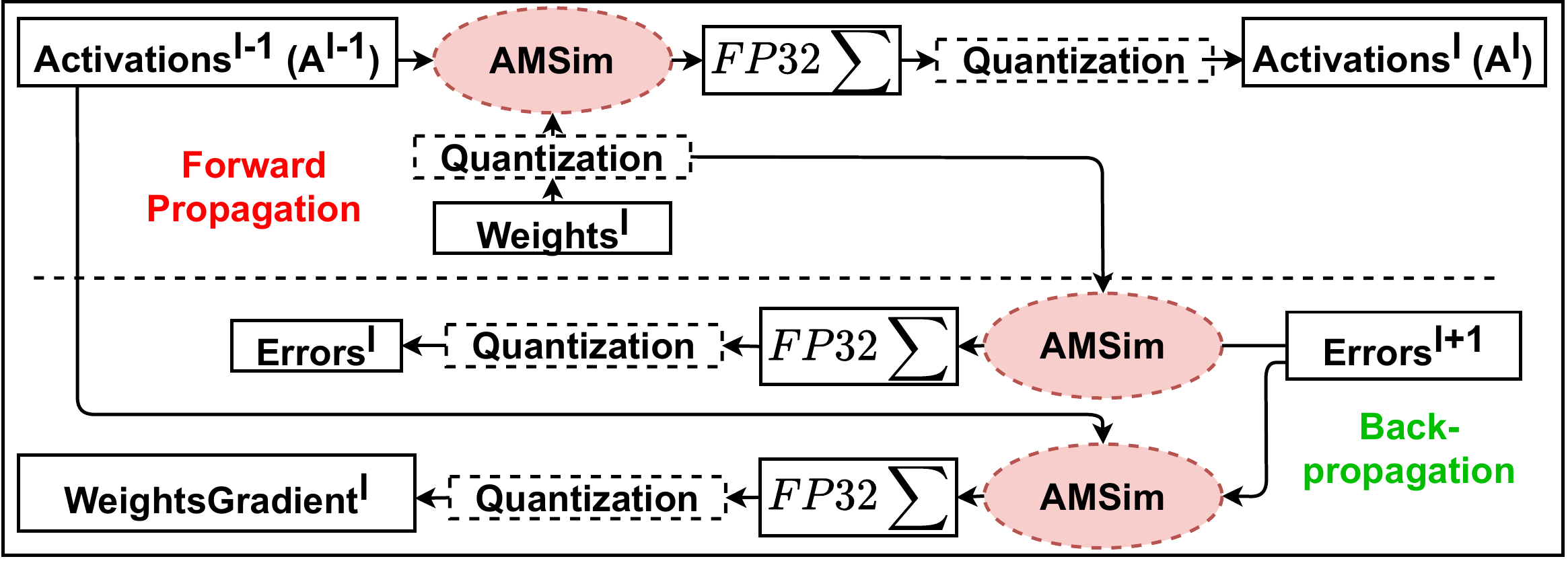}
    \caption{DNN forward and backpropagation using approximate multipliers.}
    \label{fig:mixed_precision}
\end{figure}

\section{AM Simulation: Efficient Lookup Table-based Approximate FP Multiplier Simulation} \label{sec_ams}

As explained in Section~\ref{sec_bgdnn}, a DNN consists of different types of layers, several of which involve multiplications. The computations involved in forward propagation and backpropagation of these layers were depicted graphically in Figure~\ref{fig:prop1}. To enable simulation of approximate multipliers in DNN training, all multiplications in the forward and backpropagation should be replaced by approximate multiplications, computed by AMSim as depicted in Figure~\ref{fig:mixed_precision}. \textit{However, there is no native hardware support for approximate FP multipliers on commercial processing units; hence, software simulation of approximate FP multipliers is needed.}

Efficiently simulating approximate FP multipliers in software is particularly challenging. Firstly, approximate FP multipliers have differing computation procedures, so direct C/C++ simulation (bit manipulation) cannot guarantee consistent low overhead independent of the type of approximate FP multiplier. Secondly, multiplier designers may find it challenging to optimize multipliers in order to improve simulation speed. Furthermore, GPU-based simulation must be utilized to efficiently couple approximate FP multipliers with training and inference algorithms. Therefore, this section presents a novel flow for seamlessly converting the C/C++ simulation implementation into an optimal LUT-based approximate FP simulator on GPU, AMSim. 

\begin{figure}[t]
    \centering
    \includegraphics[width=\linewidth]{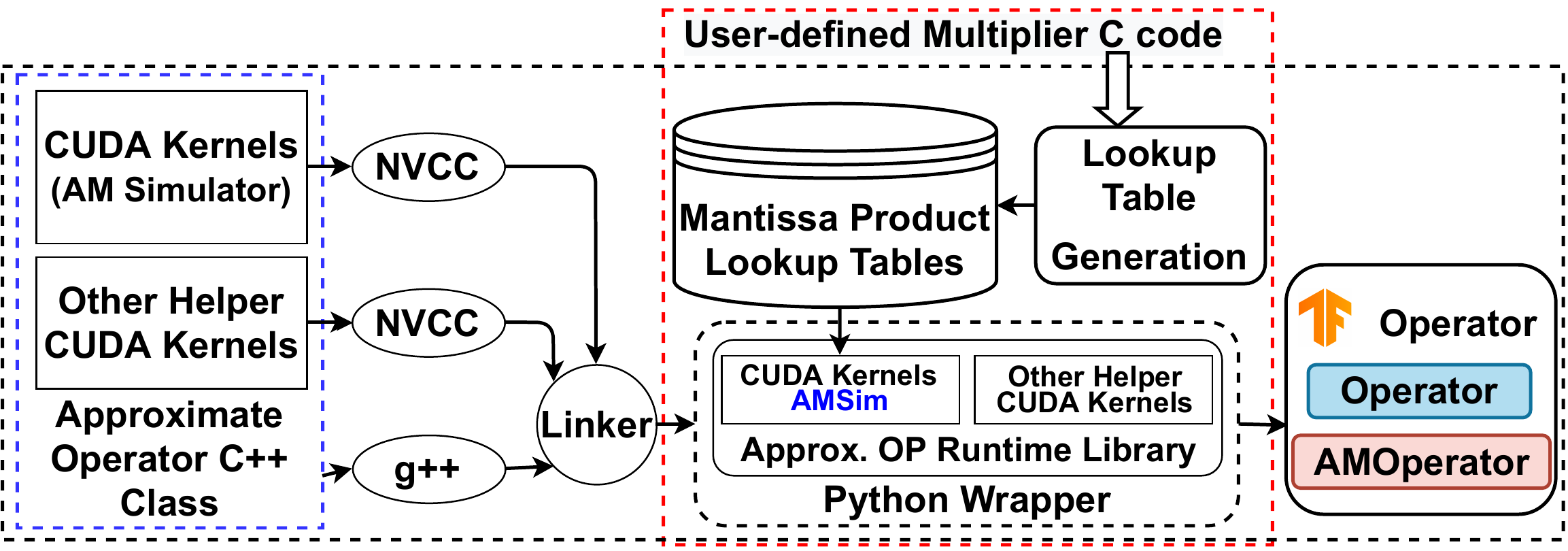}
    \caption{Overview of creation of custom ops for ApproxTrain.}
    \label{fig:gframework}
\end{figure}

As depicted in the red dash box in Figure \ref{fig:gframework}, LUT generation (see \ref{subsec_lut}) takes user-defined multiplier C/C++ code as input and generates mantissa products LUT. This generation step is required to be run once for a given approximate FP multipliers, and LUTs are written into binary files; thus, multipliers designers could load LUT binary files during run-time. Upon completing this generation step, users may load these LUTs into AMSim (see \ref{subsec_ams}) during run-time. This flow is designed based on the following key observations: (1) Mantissa multiplications contribute to 91.10\% area and 92.70\% power in the circuit of accurate FP multiplications~\cite{hassaan_2021}; thus, most AM designs~\cite{saadat_minimally_2018,saadat_realm_2020,kim_effects_2021} target optimizing the mantissa multiplications stage, and keep existing computation of exponent and sign unchanged; (2) different designs have differing approximate mantissa multiplication procedures, making it challenging to develop an efficient approximate FP multiplier simulators that will fit all designs. Mantissa multiplications are therefore simulated by LUTs (see Lookup Table Generation below), and the whole procedure for approximate FP multiplications involves three steps (see AMSim below): (1) Retrieve mantissa multiplication results from LUT; (2) Compute sign and exponent; (3) Concatenate sign, exponent, and mantissa multiplication to achieve the final approximate multiplication result.

\begin{algorithm}[!b]
\caption{Approximate Mantissa Multiplications Lookup Table Generation}
\label{algo:lutclean}
\begin{footnotesize}
\begin{flushleft}
\textbf{input: }$M$, $approx\_mul$ \Comment{\textcolor{gray}{$M\in[1, 11]$ is the bit-width of mantissa. $approx\_mul$ are approximate FP multiplication c code; it takes two FP32 numbers as inputs and outputs approximate FP multiplication as FP32.}}\\
\textbf{output: }{$mntmult\_lut$}\Comment{\textcolor{gray}{$mntmult\_lut$ is the mantissa multiplications lookup table. The size of $mntmult\_lut$ is $2^{2M}$ and each entry is 4-byte.}}\\
\end{flushleft}

\begin{algorithmic}[1]
\Function{Approximate Mantissa Multiplications Lookup Table Generation}{}
\State \textit{A $\gets$ empty FP32}; \textit{B $\gets$ empty FP32} 
\State \textit{Sign(A) $\gets$ 0 or 1}; \textit{Sign(B) $\gets$ 0 or 1} 
\State \textit{Exponent(A) $\gets$ N; Exponent(B) $\gets$ K; \newline 
\hspace*{4em}$\forall N, K \in [1, 254]$; $\forall (N+K-127) \in [1, 254]$} 
\For{\textit{k=0 to $2^M$}}
\For{\textit{j=0 to $2^M$}}
\State \textit{Mantissa(A) $\gets$ k; Mantissa(B) $\gets$ j} 
\State \textit{C $\gets$ approx\_mul($A$, $B$)}
\State \textit{un\_normalized\_exp $\gets$ Exp(A)+Exp(B)-127}
\State \textit{Carry $\gets$ 0}
\If{ $un\_normalized\_exp < Exponent(C)$ } 
\State \textit{Carry $\gets$ 1}
\EndIf
\State \textit{$mntmult\_lut[k\times2^M+j]$ $\gets$ (Carry $\ll$ 23) $\mathbin{|}$ Mantissa(C)}
\EndFor
\EndFor

\EndFunction

\end{algorithmic}
\end{footnotesize}
\end{algorithm}

\subsection{Lookup Table Generation}
\label{subsec_lut}

In~\cite{vaverka_tfapprox_2020}, authors store all approximate integer multiplication results in LUT, with the LUT occupying only 128kB of GPU memory. However, this solution is not practical for approximate FP multipliers. The de-facto industrial standard for FP is 16-bits (bfloat16 and FP16), and storing the entire result of multiplication in a LUT would require 8.6 GB of memory, \textcolor{black}{which is too costly for GPU.} We propose to store only the mantissa multiplication results in LUTs based on our above-mentioned observations. In the case of bfloat16, there are 7 mantissa bits, resulting in $2^7 \times 2^7 \times 4$ (stored as 4 bytes\footnote{Storing it as 4 bytes eliminates shift operation after retrieving from LUTs in AMSim, which further accelerates AMSim} in LUT) $= 65.53$ kB, which is negligible compared to 8 GB memory in GTX1080. \textcolor{black}{Algorithm \ref{algo:lutclean} takes the bit-width of mantissa $M$ and approximate FP multiplication C/C++ code $approx\_mul$ to generate mantissa multiplication LUTs.} In lines 2-4 of Algorithm~\ref{algo:lutclean}, two FP numbers \textit{A} and \textit{B} are initialized with arbitrary signs and exponents since the mantissa product is independent of signs and exponents. It should be noted that the exponent of \textit{A} and \textit{B} and the exponent of their product must not be special cases (0, Inf, and NaN). Otherwise, the carry from the mantissa multiplication cannot be detected; the detailed conditions are presented in line 4. In our AMSim, the carry from the mantissa multiplication is used to adjust the exponent. Lines 5-16 of Algorithm~\ref{algo:lutclean} captures the nested loop used to generate all possible mantissa combinations. The mantissa of \textit{A} and \textit{B} are populated by the nested loop indices in line 7. The populated \textit{A} and \textit{B} are then passed into the user-defined C/C++ function, \textit{approx_mul}. Then, $approx\_mul$ generates an approximate FP product \textit{C}. Lines 9-13 of Algorithm~\ref{algo:lutclean} describe how to detect carry without knowing any details about how hardware or simulations are implemented. The unnormalized exponent ($un\_normalized\_exp$) of $C$ is calculated in line 9 and is compared with the real exponent return by the user-defined C/C++ function ($approx\_mul$) in lines 11-13 to set $carry$. In AMSim, the carry bit needs to be retained in order to adjust the exponent if the real exponent of $C$ is greater than the unnormalized exponent of $C$.
Finally, carry bit and mantissa results are stored in the same entry of LUT ($mntmult\_lut$) in line 14. A script has been provided for multiplier designers to generate LUTs on the condition that a C/C++ approximate multiplication function is properly implemented by the designer.

\begin{figure}[t]
    \centering
    \includegraphics[width=\linewidth]{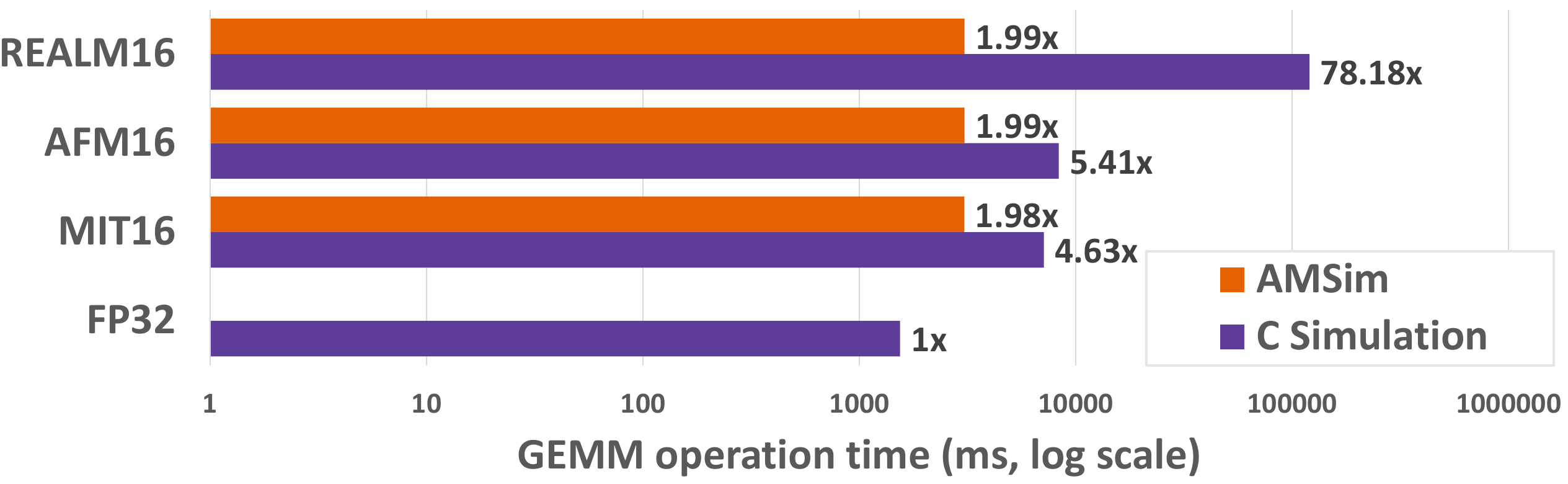}
    \caption{GEMM performance comparison for direct C Simulation and AMSim for multipliers REALM16~\cite{saadat_realm_2020}, AFM16~\cite{saadat_minimally_2018} and MIT16~\cite{mitchell_computer_1962}. Note, FP32 is the time used by native hardware.}
    \label{fig:GEMMAMS}
\end{figure}

\subsection{AMSim}
\begin{algorithm}[!b]
\caption{Approximate FP Multiplication Simulator (AMSim)}
\label{algo:AM}
\begin{footnotesize}
\begin{flushleft}
\textbf{input: }$a$, $b$, $mntmult\_lut$ \Comment{\textcolor{gray}{$a$ and $b$ are FP inputs to the simulation. $mntmult\_lut$ is the mantissa product lookup table}}\\
\textbf{output: }{$c$}\Comment{\textcolor{gray}{$c$ is the approximate product of $a$ and $b$}}\\
\end{flushleft}

\begin{algorithmic}[1]
\Variables
 \State \textit{M}, Mantissa Bit-width.
 \State \textit{M\_MASK}, Mantissa Mask.
 \State \textit{E\_MASK}, Exponent Mask.
\EndVariables
\Function{Approximate FP Multiplication Simulation}{}

\State \textit{Amnt $\gets $ M_MASK $\&$ a};  \textit{Bmnt $\gets $ M_MASK $\&$ b} 
\State \textit{Mntmult$\gets$mntmult_lut[Amnt $\gg (23-M\times2)$+\newline
\hspace*{10.5em}Bmnt $\gg (23-M)$]} 
\State \textit{Carry $\gets$ Mntmult $\&$ 0x00800000} 
\State \textit{Mntmult$\gets$ Mntmult $\&$ 0xFF7FFFFF}
\State \textit{Sign$\gets$ (a $\oplus$ b) $\&$ S_MASK}
\State \textit{Exp$\gets$((a $\&$ E_MASK + b $\&$ E_MASK) $\gg$ 23) - 127}
\If{\textit{ $Exp \le 0$ or a $\&$ E\_MASK == 0 or b $\&$ E\_MASK == 0}} 
\State \textit{c $\gets$ 0}
\ElsIf{$Exp \ge 255$} 
\State \textit{c $\gets$ INFINITY}
\Else 
\State \textit{Exp $\gets$ Exp + Carry}
\State \textit{$c\gets Sign \mathbin{|} (Exp \ll 23) \mathbin{|} Mntmult$}
\EndIf
\EndFunction

\end{algorithmic}
\end{footnotesize}
\end{algorithm}
\label{subsec_ams}
\textcolor{black}{The AMSim is proposed to simulate approximate FP multipliers on GPU; it is composed of the three steps mentioned earlier in section \ref{sec_ams}, and Algorithm \ref{algo:AM} elaborates this mechanism in detail. Algorithm \ref{algo:AM} takes two FP numbers $a$, $b$ and mantissa product LUT; it output the approximate product of $a$ and $b$.} In line 7, the mantissa of $A$ and $B$ are extracted; then, on line 8, the index to fetch LUT is computed by concatenating the mantissa of $A$ and $B$. In lines 9-10 of Algorithm~\ref{algo:AM}, the mantissa multiplication results and carry are decoupled. In line 10, the sign of the approximate multiplication output $C$ is computed as the XOR (exclusive-or) of the signs of $A$ and $B$. Exceptional cases (0 and Infinity) and normal cases are handled from lines 11 to 17 of Algorithm~\ref{algo:AM}. If either the biased exponent of $C$ is not greater than zero or one of the inputs ($A$ or $B$ is zero) (line 12), then $C$ should be zero. When the biased exponent of $C$ exceeds or equals 255, the $C$ is overflowed and results in Infinity. In lines 16-18 of Algorithm~\ref{algo:AM}, the biased exponent is adjusted based on $carry$ in the normal case. As a final step, sign, exponent, and mantissa are concatenated to form $C$. The AMSim is implemented as an inline device function and compiled into a part of the CUDA kernel. The LUT is retrieved from texture memory \textcolor{black}{on GPU}, a similar approach to that described in work~\cite{vaverka_tfapprox_2020}. Texture memory has its dedicated texture cache, connected to the L2 cache, and it would not affect the DNN workload in the L1 cache; thus, this approach reduces memory transaction overhead.
\textit{Note that, despite giving 16-bit FP as an example, our approach enables generic $(1, e, m)$ FP approximate multiplication simulation; bits of mantissa $m$ could be selected from 1 (16 Bytes) to 11 (16.8MB, 1.6\% of total GTX1080 memory), supporting a wide range of precisions. \textcolor{black}{Additionally, the bits of the exponent $e$ can be varied from 1 to 8 provided that a proper exponent casting function is given.}} 

\subsection{GEMM performance evaluation}

As shown in Figure \ref{fig:GEMMAMS}, we evaluate AMSim and direct C simulation with different approximate multipliers in GEMM (general matrix multiplication, see section \ref{sec_customcuda}). The two input matrices to GEMM are both 8000 by 8000, and the experiment is performed using GTX1080 GPU. Our approach, AMSim, is consistently 2x slower than native hardware FP32 for REALM16, AFM16, and MIT16, while other direct C simulations have performance overhead between 4.6x and 78.2x.

In the following section, we present the integration of the proposed AMSim into the framework, ApproxTrain, to enable DNN training/inference using approximate FP Multipliers.
\section{ApproxTrain}
\label{sec_approxtrain}
\begin{figure}[t]
    \centering
    \includegraphics[width=\linewidth]{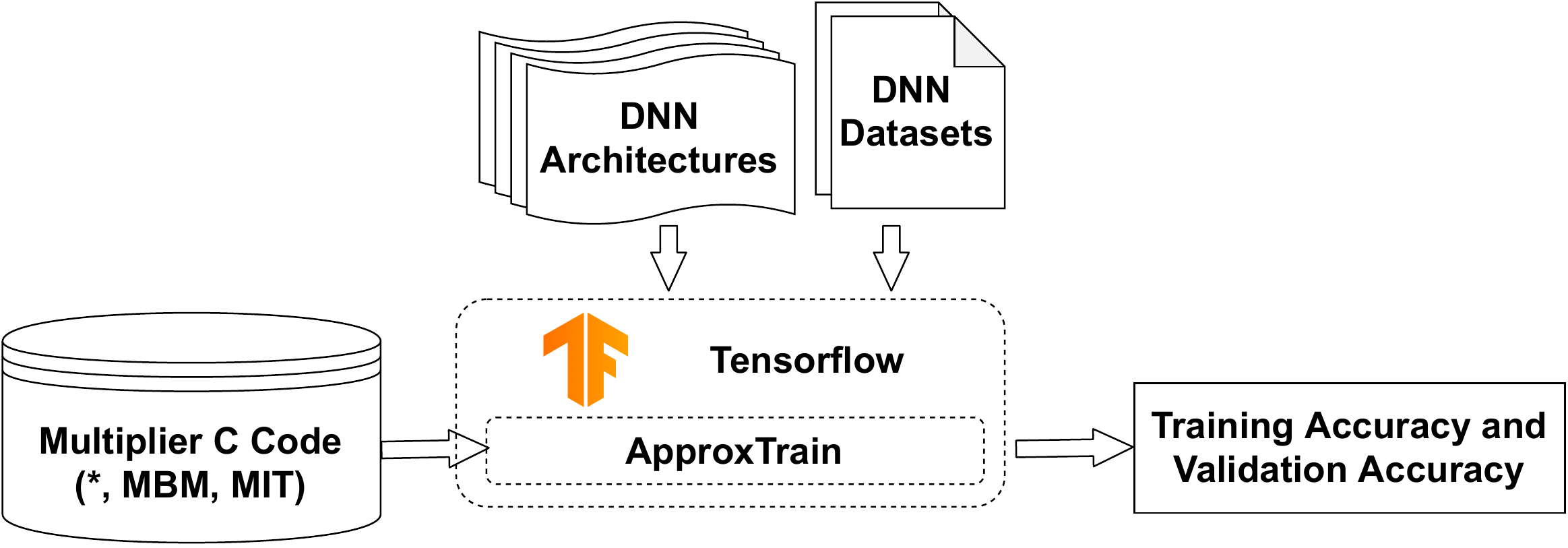}
    \caption{ApproxTrain: From end-user Perspective.}
    \label{fig:enduserperspective}
\end{figure}

ApproxTrain integrates AMSim into TensorFlow, so that different DNN architectures can be efficiently constructed and evaluated using high-level APIs. In ApproxTrain, we create custom TF \textit{ops} (see Section~\ref{sec_background} for 
explanation of \textit{ops}) to support different types of DNN layers with approximate multiplication. To equip our custom \textit{ops} with AMSim, we developed GPU-accelerated custom CUDA kernels for the implementation of our custom TF \textit{ops}. 
These custom CUDA kernels are needed because the standard \textit{ops} available in the open-source TF library use closed-source cuDNN/cuBLAS libraries in the backend that cannot be modified. Thus, ApproxTrain enables fast evaluation of training/inference of different DNN architectures using different approximate multiplier designs.

In the following subsections, we first present an overview of ApproxTrain, followed by a detailed description of our approach to create two custom TF \textit{ops} and the underlying CUDA kernels.

\subsection{ApproxTrain: Framework Overview}



An overview of the ApproxTrain use-case is shown in Figure~\ref{fig:enduserperspective}. In addition to the normal design flow of TF, a user simply needs to: (1)  provide functional models of the approximate multipliers in C/C++; (2) and replace the standard layer \textit{ops} with the approximate versions from ApproxTrain in the DNN architecture. Example code snippets for such a replacement is demonstrated in Listing1 and Listing2. After importing the compiled library of the custom \textit{ops} from ApproxTrain, the DENSE \textit{op} (fully connected) and CONV2D \textit{op} (convolutional layer) are simply replaced with their approximate versions AMDENSE and AMCONV2D, respectively.

\smallskip

\lstset{language=Python}

\begin{lstlisting}[ caption={DNN model using standard TensorFlow op for convolutional and dense layer.},
captionpos=b, ,style=base]
from tensorflow.keras import layers

model = models.Sequential()
model.add(layers.Conv2D(32, (3, 3))
model.add(layers.Dense(32, (3, 3))
\end{lstlisting}

\begin{lstlisting} [caption={DNN model using ApproxTrain for approximate convolutional and dense layers.},
captionpos=b,style=base]
from python.keras.layers.am_convolutional import @AMCONV2D@
from python.keras.layers.am_dense import @AMDENSE@

model = models.Sequential()
model.add(layers.@AMCONV2D@(32, (3, 3))
model.add(layers.@AMDENSE@(32, (3, 3))
\end{lstlisting}


Figure \ref{fig:gframework} \textcolor{black}{depicts
an} overview of the internals of creation and compilation of our custom \textit{ops} in ApproxTrain. The main component is the
approximate operator C++ class inside the blue dashed box which has multiple operations such as input validation, serializing tensors to linear arrays, memory allocation, and performing computations. The computational part of approximate operator C++ class includes functions to calculate feedforward propagation and back propagation by invoking our custom CUDA kernels or CPU kernels\footnote{The CPU implementation was used for validating our GPU implementation and benchmark, but could also be used by a user who does not have GPU access at the cost of higher run-time. 
}. Custom CUDA kernels (explained in Section~\ref{sec_customcuda}) are responsible for linear algebra operations and data rearrangement and are equipped to use AMSim. The AMSim is implemented as a device function for running on GPU. As stated before, custom CUDA kernels are written from scratch because the closed-source cuDNN and cuBLAS libraries cannot be modified to use approximate multipliers.

All CUDA kernels are compiled by NVCC, and the C++ operator class is compiled with g++. Then, the compiled C++ object files are linked with the complied CUDA kernel objects to form the approximate operator shared library. This approximate operator run-time library is then enclosed in a python wrapper which is then registered into the standard TF library. Note that the compilation steps above only need to be done once.
Instead of replacing the corresponding original operators in TensorFlow, the new approximate operators are kept alongside the original ones. Given user-defined approximate multiplier C codes, LUTs can be obtained by \textit{Lookup Table Generation} (explained in \ref{subsec_lut}), as depicted in Figure \ref{fig:gframework}. The obtained LUTs are loaded into the approximate operator run-time library during run-time to simulate different functional models of approximate multipliers. These python wrappers have the same parameters as original operators, and the users simply need to change the name of the original operators to the approximate ones to simulate the approximate multipliers, as demonstrated in the code listings above.

We have currently added two operators, AMDENSE (approximate Dense layers) and AMCONV2D (approximate Conv2D layers), to our framework to enable the support of two layers: Dense and the Conv2D layers. These two operators allow us to cover a large portion of DNN architectures. 

\begin{figure*}[!t]
 \begin{center}

\resizebox{\linewidth}{!}{\includegraphics{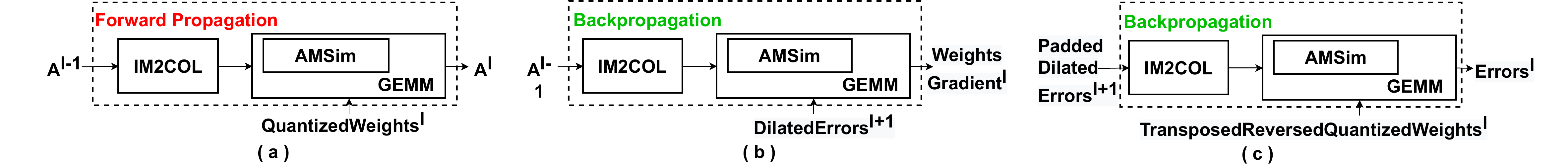}} 

\end{center}

\caption{Forward propagation and back-propagation implementation overview.}
 \label{fbpgpu1}
\end{figure*}

\subsection{AMCONV2D: Conv. Layer Custom Op with Approximation}

This subsection explains the approach used to realize the forward and backward propagation in the custom approximate operator AMCONV2D.
\begin{algorithm}[!b]
\caption{Approximate Forward propagation}
\label{algo:forward}
\begin{footnotesize}
\begin{flushleft}
\textbf{input: }$A^{l-1}$,  $W^l$, $S$, $P$, $LUT$ \Comment{\textcolor{gray}{$A^{l-1}$: activation from layer $l-1$; $W^l$: weight from layer $l$; $S$: stride; $P$: Padding; $LUT$: mantissa product LUT}}\\
\textbf{output: }{$A^{l}$} \Comment{\textcolor{gray}{$A^{l}$: activation from layer $l$}}\\
\end{flushleft}

\begin{algorithmic}[1]
\Function{Approximate Forward Propagation}{}

\State \textit{PSize},\textit{ColSize} $\gets$ \textit{calculate\_size}s($A^{l-1}$,$W^l$,$P$,$S$) \Comment  \textcolor{gray}{$PSize$: The size of padding $ColSize$: The size of Im2Col results}
\State \textit{allocate_GPU_memory}(\textit{PSize},\textit{ColSize},\textit{LUT}) \Comment  \textcolor{gray}{For $A^{l-1}$, \textit{Columns}(output of IM2COL), $W^{l}$, $a^l$ and $LUT$. }

\State  \textit{Columns $\leftarrow$ IM2COL_kernel($A^{l-1}$, PSize, ColSize)}\;
\State  \textit{$A^l$ $\leftarrow$ GEMM_kernel(Columns, $W^l$, $LUT$ ...)}\Comment \textcolor{gray}{... refer to \textit{m,n,k,lda,ldb and ldc} that are ommited for simplicity. $A^l$ is the output activation}

\EndFunction

\end{algorithmic}
\end{footnotesize}
\end{algorithm}

\textbf{Forward propagation:} Forward propagation takes activations $A^{l-1}$ and $QuantizedWeights^l$ to compute output activations $A^{l}$ (Figure \ref{fig:prop1}). We use the IM2COL+GEMM approach~\cite{chellapilla_high_nodate,chetlur_cudnn_2014} to perform forward propagation, because this approach exposes fine-grained parallelism suitable for GPU acceleration. Figure \ref{fbpgpu1}~(a) illustrates this approach. In Figure \ref{fbpgpu1}~(a) $A^{l-1}$ is the input to IM2COL operation. The output of IM2COL and $QuantizedWeights^l$ are subjected to the GEMM operation. The GEMM kernel contains AMSim that can invoke native hardware multiplications (* operator) or perform approximate multiplications using LUTs. Algorithm \ref{algo:forward} describes our approach for forward propagation on GPU. First, the sizes of GPU global memory arrays are computed (line 2 in Algorithm \ref{algo:forward}) and allocated (line 3). Then, the IM2COL Kernel is invoked on the GPU (line 4 in Algorithm \ref{algo:forward}), followed by the GEMM kernel (line 5). Details of these GPU kernels will be discussed later in Section \ref{sec_customcuda}.

Despite not being shown in Algorithm \ref{algo:forward} for simplicity, a loop that invokes the kernels iteratively on tiles of the array $A^{l-1}$ is implemented to enable our framework to train large architectures and datasets. This is because the CUDA grid (group of blocks of CUDA threads) dimension along the y-axis is limited to 65535 \cite{noauthor_cuda_nodate}, and thus large input data cannot be fit into the GPU grid entirely.

\smallskip

\textbf{Backpropagation:}
The backpropagation involves two gradient computations: weights gradient and gradient for the preceding layer (labeled as $Errors^{l}$ in Figure~\ref{fig:prop1}). 
Algorithm \ref{algo:backward} elaborates our backpropagation approach for efficient GPU implementation. Similar to Algorithm \ref{algo:forward}, here too, we first calculate the GPU array sizes and allocate them in lines 2-3 of Algorithm \ref{algo:backward}. Lines 4-5 are the invocation of the kernels for the weights gradient (explained below), and lines 6-8 are for the preceding layer gradient (explained below). Note that for backpropagation, we also implemented tiling (as we have explained for forward propagation) despite not being shown in Algorithm \ref{algo:backward}. 

\subsubsection{Weights Gradient Computation}

We restructured the weights gradient computation to exploit the IM2COL+GEMM approach as illustrated in Figure \ref{fbpgpu1}~(b). We first subject $Errors^{l+1}$ to dilation (inserting zeros between elements based on the $stride$ parameter). Then, this $DilatedErrors^{l+1}$ is fed to GEMM (Figure \ref{fbpgpu1}~(b)).

\begin{algorithm}[!b]
\caption{Approximate  Backpropagation}
\label{algo:backward}
\begin{footnotesize}
\begin{flushleft}
\textbf{input: }$a^{l-1}$, $W^l$, $Error^{l+1}$, $Stride$ \Comment{\textcolor{gray}{$a^{l-1}$ is the activation from layer $l-1$),$W^l$ is the weight from layer $l$}}\\
\textbf{output: }{$W^{l\prime}$, $Errors^{l}$}\Comment{\textcolor{gray}{$W^{l\prime}$ is the gradient of $W^l$}}\\
\end{flushleft}

\begin{algorithmic}[1]
\Function{Approximate Backpropagation}{}

\State \textit{PSize, ColSize $\gets$ calcualte\_sizes($a^{l-1}$, $W^l$, $Errors^{l+1}$)}\Comment{\textcolor{gray}{$PSize$: The size of padding $ColSize$: The size of Im2Col results}} 
\State \textit{allocate_GPU_memory(PSize, ColSize, LUT)} \Comment \textcolor{gray}{For $a^{l-1}$, Columns, $DilatedError^{l+1}$, $W^{l\prime}$ and $Errors^{l}$}
\State \textit{$Columns_{a^{l-1}}$ $\gets$ IM2COL_Weight_Kernel($a^{l-1}$, ColSize, PSize)}\;
\State \textit{$W^{l\prime}$ $\gets$ GEMM_Kernel($Columns_{a^{l-1}}$, $Error^{l+1}$, $LUT$ ...)}

\State \textit{$Columns_{PDError^{l+1}}$ $\gets$ IM2COL_PLG_Kernel($Error^{l+1}$, ColSize, PSize)}\Comment{\textcolor{gray}{\textit{IM2COL kernel for preceding layer gradient (PLG)}, \textit{$PDError^{l+1}$} is the $PaddedDilatedError^{l+1}$}}
\State \textit{$(W^l)^T_r$ $\gets$ Reverse_Transpose_kernel($W^l$)}
\State \textit{$Errors^{l}$ $\gets$ GEMM_Kernel($Columns_{PDError^{l+1}}$, $(W^l)^T_r$, $LUT$...)}

\EndFunction

\end{algorithmic}
\end{footnotesize}
\end{algorithm}

As opposed to forward propagation, mapping backpropagation to GEMM along with IM2COL to efficiently exploit GPU architecture is challenging. A naive method to implement the mapping of computation of weights gradient to GEMM would be to implement a separate GPU kernel to perform the dilation operation and invoke it before the GEMM kernel. However, this naive method would be inefficient due to two reasons. First, invoking a kernel unnecessarily adds extra performance overhead. Second, a dilated array would require several times the memory as the original array (depending on \textit{stride} value), consequently reducing the number of non-zero elements that can be stored in the GPU global memory, thus requiring more tiling (similar to tiling explained in forward propagation above). Instead of such a native approach, we implicitly perform this dilation inside the IM2COL\_Weight\_Kernel (a modified IM2COL kernel) by skipping elements in $A^{l-1}$ that correspond to zero (line 4 of Algorithm \ref{algo:backward}) if the $Error^{l+1}$ array was dilated.  

\subsubsection{Preceding Layer Gradient}

We also restructure the computation of the preceding layer gradient to exploit the IM2COL+GEMM approach as shown in Figure \ref{fbpgpu1}~(c). For this, we first subject $Errors^{l+1}$ to dilation (inserting zeros between elements as explained before), followed by padding (inserting zeros around the image along height and width dimensions). This $PaddedDilatedErrors^{l+1}$ is the input to IM2COL as shown in Figure \ref{fbpgpu1}~(c).  Then, we subject $QuantizedWeights^l$ to transposition and reversal of elements. This $TransposedReversedQuantizaedweights^{l}$ is fed as the input to the GEMM operation as shown in Figure~\ref{fbpgpu1}~(c) along with the output of IM2COL.

Exploiting GEMM in preceding layer gradient ($Errors^{l}$ in Figure \ref{fbpgpu1}~(c)) of AMCONV2D for efficient execution on the GPU is even more non-trivial since both transposition and reversal of elements in $Weights^l$ are involved, in addition to padding and dilation of $Errors^{l+1}$. 

Instead of having a separate kernel for dilating $Errors^{l+1}$ which would cause additional kernel invocation overhead, we integrate the dilation operation into the IM2COL\_PLG\_Kernel (a modified IM2COL Kernel that performs padding and dilation) where each thread copies a zero into IM2COL results if the current pixel is at a dilated position.
Unfortunately, unlike in backpropagation for weights gradient where the second operand to GEMM requires dilation, the dilation must be performed on the input to the IM2COL. Thus, we cannot simply skip elements as is done for weights gradient computation despite the need for more GPU memory.

Transposition and the reversal of $QuantizedWeights^l$ can be implicitly done inside the GEMM kernel by manipulating the array index when accessing the second operand for GEMM. However, this would be highly inefficient because the global memory access pattern would not enable memory coalescing. Thus, here it is better to sacrifice some time to invoke a separate kernel that solely performs the reversal and transposition of $QuantizedWeights^l$, so that more time can be saved during the memory accesses of GEMM operation. 

Since AMCONV2D is implemented by the GEMM approach, all multiplications are done in GEMM kernel; thus, we replace accurate multiplication in GEMM with AMSim device function to enable simulation.

\subsection{AMDENSE: Dense Layer Custom Op with Approximation}

Unlike in the convolution layers, in the dense layer, each neuron receives input from all neurons in the preceding layer (see Figure \ref{fig:denseiluu}). Like AMCONV2D described above, forward propagation and backpropagation of AMDENSE need to be implemented to realize training. Compared to AMCONV2D, AMDENSE occurs in a small proportion of the total computation and thus contributes to a tiny fraction of the total training time. 
Thus, CUDA optimization efforts are not as crucial as for AMCONV2D. 

\smallskip
\textbf{Forward propagation:}
Forward propagation can be mapped to a matrix-vector multiplication where weights in the dense layer are a 2-dimensional matrix, and the activations from the preceding layer are a 1-dimensional vector. This is shown using a simplified example in Figure \ref{fig:denseiluu} where the dense layer output is computed as:
$\big(\begin{smallmatrix}
  o_1\\
  o_2
\end{smallmatrix}\big)=\big(\begin{smallmatrix}
  w_{11} & w_{12} & w_{13}\\
  w_{21} & w_{22} & w_{23}
\end{smallmatrix}\big) 
\big(\begin{smallmatrix}
  x_1\\
  x_2 \\
  x_3
\end{smallmatrix}\big) = \big(\begin{smallmatrix}
  w_{11}x_1 + w_{12}x_2 + w_{13}x_3\\
  w_{21}x_1 + w_{22}x_2 + w_{23}x_3
\end{smallmatrix}\big)$; where $x$ is the activations, $w$ is the weights. We implemented a separate matrix-vector multiplication CUDA kernel for this rather than using the previously used GEMM kernel, because shared memory-based tiling is superfluous for a 1-D vector.

\smallskip
\textbf{Backpropagation:}
Similar to AMCONV2D, backpropagation in AMDENSE also involves computations of weights gradient and preceding layer gradient.

\subsubsection{Weights Gradient Computations}

The gradient of weights in the AMDENSE layer $l$ is computed as $\delta_{out}a_{in}^{T}$ where $a_{in}$ is the activation from preceding layer $l-1$ and $\delta_{out}$ is the error backpropagated from succeeding layer $l+1$. The gradient of weight in Figure \ref{fig:denseiluu} is computed as $\big(\begin{smallmatrix}
  w_{11}^{'} & w_{12}^{'} & w_{13}^{'}\\
  w_{21}^{'} & w_{22}^{'} & w_{23}^{'}
\end{smallmatrix}\big)=\big(\begin{smallmatrix}
  o_1^{'}\\
  o_2^{'}
\end{smallmatrix}\big)\big(\begin{smallmatrix}
  x_1 & x_2 & x_3
\end{smallmatrix}\big)$. 
The same matrix-vector multiplication CUDA kernel is used here.

\subsubsection{Preceding Layer Gradient Computations}
\begin{figure}[t]
    \centering
    \includegraphics[width=1.0\linewidth]{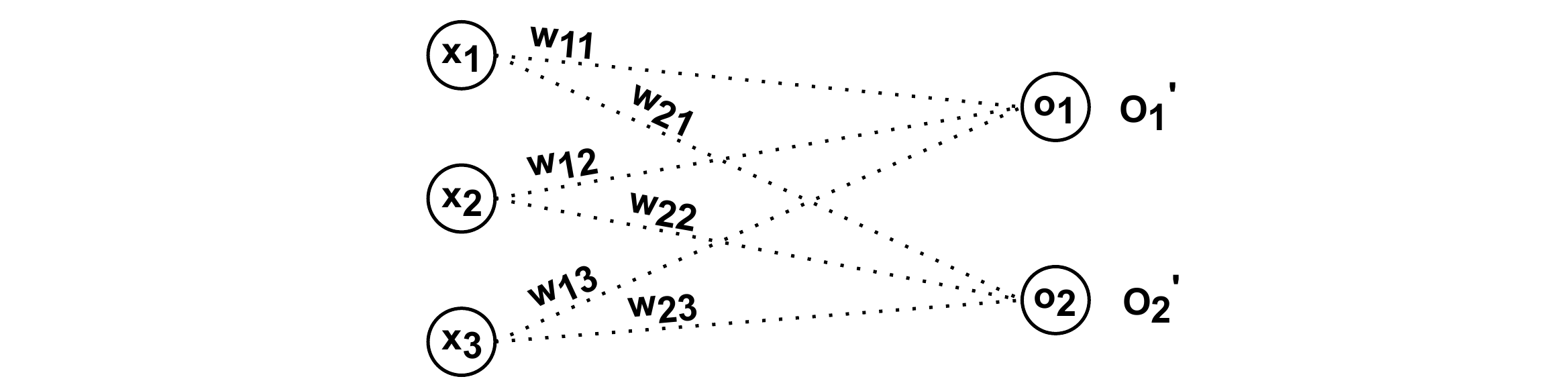}
    \caption{AMDENSE implementation illustration.    \label{fig:denseiluu}}
\end{figure}

The gradient of input in the AMDENSE layer $l$ is calculated as $(w)^T\delta_{out}$ where $(w)^T$ is the transpose of the weights in the layer $l$. The gradient of input of given example in Figure \ref{fig:denseiluu} can be computed as $\big(\begin{smallmatrix}
  x_1^{'}\\
  x_2^{'} \\
  x_3^{'}
\end{smallmatrix}\big)=\big(\begin{smallmatrix}
  w_{11} & w_{21}\\
 w_{12} & w_{22}\\
 w_{13} & w_{23}
\end{smallmatrix}\big) \big(\begin{smallmatrix}
  o_1^{'}\\
  o_2^{'}
\end{smallmatrix}\big)$.
For the computation of the preceding layer gradient, we use the same matrix-vector kernel used for forward propagation. The transposition of the vector is implicitly handled because the elements are anyway stored linearly in memory.

We replace accurate multiplications in matrix-vector kernel with AMSim, considering that the matrix-vector kernel contains all multiplications of AMDENSE.

\subsection{Other Custom CUDA Kernels}
\label{sec_customcuda}

The custom AM ops described above utilize several custom CUDA kernels. We developed these kernels to replace the kernels offered by the closed-source cuDNN and cuBLAS library. These custom CUDA kernels (which involve multiplication) are equipped to use AMSim to perform multiplication. In simple terms, these kernels may call the approximate multiplier functions with two operands as the arguments instead of using the `*' operator to multiply the two operands.
A brief description of these custom kernels is given below.

\smallskip
\textbf{GEMM kernel:}
The GEMM kernel is a highly optimized kernel that uses a 2-D threading indexing model with 16x16 as the CUDA thread block size. 16x16 tiles of the input matrices are fetched to fast GPU shared memory (on-chip SRAM) from global memory to be used for repeated memory accesses. 

\smallskip
\textbf{IM2COL kernels:} 
There are three separate IM2COL kernels, as we mentioned before: 1. IM2COL (line 4 of the Algorithm \ref{algo:forward}) for forward propagation 2. IM2COL\_Weight\_Kernel (line 4 of the Algorithm \ref{algo:backward}) for weights gradient and 3. IM2COL\_PLG\_Kernel (line 6 of the Algorithm \ref{algo:backward}) for preceding layer gradient. IM2COLs mentioned above are implemented by utilizing a 1-D threading indexing model with 256 as the CUDA thread block size.

\textit{IM2COL:}
Each thread in IM2COL first locates the element position of $A^l$ (the output of forward propagation), then locates the patch's element position (a flattened window) corresponding to $A^l$. The above two steps are needed to copy input data into the correct output position. Finally, the element in the input is located and copied to the IM2COL output.

\textit{IM2COL\_Weight\_Kernel:}
The IM2COL\_Weight\_Kernel first locates the element position of $WeightsGradient^l$ rather than $A^l$ in forward propagation since its output is the $WeightsGradient^l$. Then, the IM2COL\_Weight\_Kernel locates the element position in the patch related to $WeightsGradient^l$. Finally, the IM2COL\_Weight\_Kernel locates the element in $A^{l-1}$ and copies it to the IM2COL output; note that skipping elements is performed here if $stride$ is greater than 1.

\textit{IM2COL\_PLG\_Kernel:}
Similar to IM2COL and IM2COL\_PLG\_Kernel, IM2COL\_PLG\_Kernel first locates the element position of the preceding layer gradient ($Errors^{l-1}$ in Figure \ref{fbpgpu1}~(c) and then locates the element position in the patch. After the above two steps, the element position of input is located. However, this element position of input is  computed based on the size of $PaddedDilatedErrors^{l+1}$ rather than $Errors^{l+1}$ (note the input data to IM2COL\_PLG\_Kernel is still $Errors^{l+1}$, but the size of input data is set to that of the $PaddedDilatedErrors^{l+1}$); thus, an additional procedure is implemented for each thread to check if current position is dilated position or not. The native IM2COL could handle padding, but the computation for the size of padding is different from forward propagation and weights gradient, despite not being explained here.

\smallskip
\textbf{Transpose-And-Reverse Kernel:}
The TransposeAndReverse Kernel is a custom CUDA kernel that swaps dimensions of data and reverses elements order. It uses a 2-D threading indexing model with 32x32 as the CUDA block sizes. It first gets the index of a pixel along the height and width dimension of input to reverse the elements. Then, it gets the index of the dimension that is to be swapped in the following procedure. Then, this kernel starts swapping and reversing elements by manipulating the index. This kernel improves spatial locality by rearranging data order; thus, when GEMM kernel loads data into shared memory, memory coalescing occurs.

\smallskip
\textbf{Matrix-Vector Multiplication Kernel:}

Matrix-vector multiplication custom CUDA kernel is implemented by 1-D threading mode with 1024 threads in each block. Each thread will operate multiplication $n$ times ($n$ depends on the length of the vector).



\section{Experiment Setup}
\label{sec_expsetup}
We use the presented ApproxTrain to perform a series of \textit{training} and \textit{inference} experiments for image classification. The experiments uses various DNN architectures and datasets on different platforms. 
The purpose of these experimental evaluations is two-fold: (1) evaluate the efficacy of approximate multipliers in DNN training, i.e., the effect on training convergence and accuracy (Section~\ref{sec_results_accuracy}); and (2) evaluate the timing performance of ApproxTrain with different DNN architectures/datasets/platforms (Section~\ref{sec_results_performance}).
In the experiments, the framework inputs are neural network architecture, dataset, and multiplier-type. The outputs are the timing performance numbers and validation and test accuracy of the classification task.
The details of different datasets, neural network architectures, and other settings used in our experiments are below.

\textbf{Datasets:} Three popular datasets from image-classification are used in our experiments: MNIST \cite{deng_mnist_2012}, CIFAR-10 \cite{krizhevsky_learning_nodate} and ImageNet \cite{deng_imagenet_2009}. MNIST is hand-written digits consisting of 60,000 training and 10,000 test samples. Each sample is a $32{\times}32$ gray image. In the CIFAR10 dataset, there are ten classes with 6000 $32{\times}32$ coloured images per class. The dataset is divided into 50,000 training samples and 10,000 test samples. 
ImageNet contains 1.2 million training images, spanning 1000 object classes. MNIST and CIFAR-10 are usually considered small datasets, while ImageNet is one of the largest datasets available for image classification.

\textbf{Neural Network Architectures:} Five neural network architectures: LeNet-300-100, LeNet-5 ~\cite{lecun_gradient-based_1998}, ResNet-18/34/50~\cite{he_deep_2016} are used in our experiments.
LeNet-300-100 is a multi-layer perceptron (MLP), while LeNet-5 is a convolutional neural network (CNN) having two convolution layers and three dense layers. 
ResNet is a deep convolutional network whose complexity can be adjusted by adding or removing building blocks~\cite{he_deep_2016}. \textcolor{black}{ResNet-18/34/50 contains 18, 34 and 50 layers,  respectively.}
The various combinations of datasets and architectures used in our experiments are listed in the first \& second columns of Table~\ref{tab:finalacc1}.

\textbf{Datatype:}
Our experiments use floating-point format instead of integer/fixed-point because training typically requires a higher dynamic range. We keep the sign \textit{(s)} as 1-bit and the exponent \textit{(e)} as 8-bit (similar to FP32 and bfloat16~\cite{wang_bfloat16_2019}). The number of mantissa bits \textit{(m)} is varied in different experiments to achieve different bit-widths. The details of various data types/multipliers and their bit widths are listed in Table~\ref{tab_dtandmul}. 
Since exponents are the same in all formats, type-conversion is simply a matter of bit-truncation or bit-extension. All accumulation operations are performed in FP32 to realize the industry de-facto standard of mixed-precision training when lower bit-widths are used for multiplication~\cite{amulya_vishwanath_video_2019}.

\textbf{Experiment platforms:}

 \textcolor{black}{Three types of platforms are used for the experiments. System-I is equipped with a single NVIDIA V100 GPU and 12 core Intel Xeon Scalable (Cascade Lake) processor (24 CPUs per core). 
 System-II is equipped with GTX1080 and i7 6600 CPU. System-I and System-II are used for run-time performance benchmarking (Section~\ref{sec_results_performance}).
 In addition, to run the training convergence test for large datasets (Section~\ref{sec_results_accuracy}), we used another system which is a two-node cluster equipped with 8 V100 GPUs and two full Intel Xeon Scalable 'Cascade Lake' cores.  To realize multi-GPU (distributed) training environment on the high-end cluster, TensorFlow wrapped by Horovod is used. The operating system used on all systems is Ubuntu 18.04. }

\textbf{Implementation Details:}
\textcolor{black}{The presented framework is integrated into TensorFlow 2. The tested TensorFlow version is 2.3.0, which requires CUDA 10.1 and cuDNN 7.6.5. The custom CUDA kernels are compiled with NVCC provided by CUDA 10.1, whereas the supporting C/C++ files are complied with gcc-8/g++-8.}


\begin{figure*}[!t]

 \begin{center}

 \begin{tabular}{ c c c }
{ (a) MNIST/LeNet-300-100 } &  { (b) MNIST/LeNet-5 }  & { (c) CIFAR-10/ResNet18 }\\
\resizebox{0.29\textwidth}{!}{\includegraphics{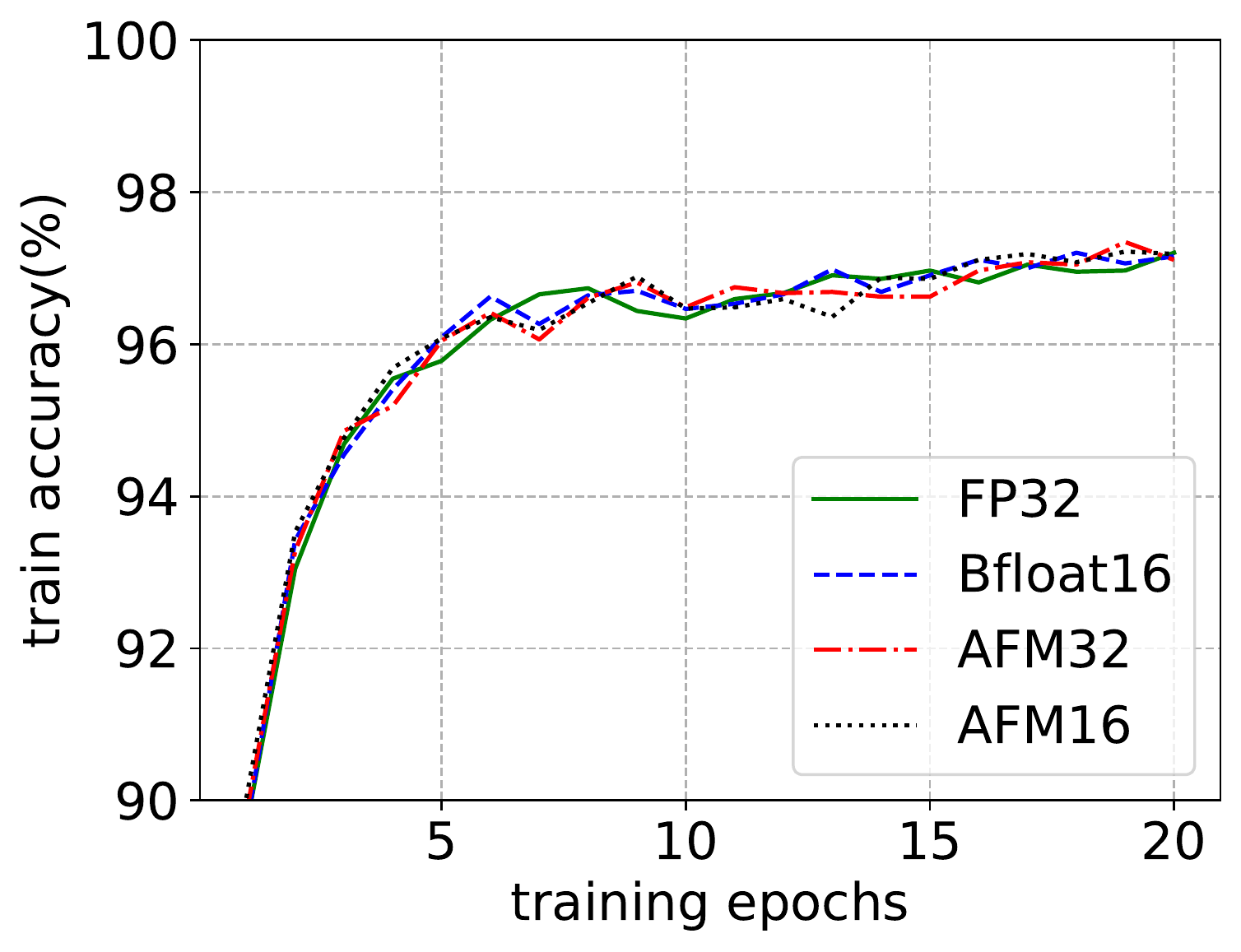}}  &  \resizebox{0.29\textwidth}{!}{\includegraphics{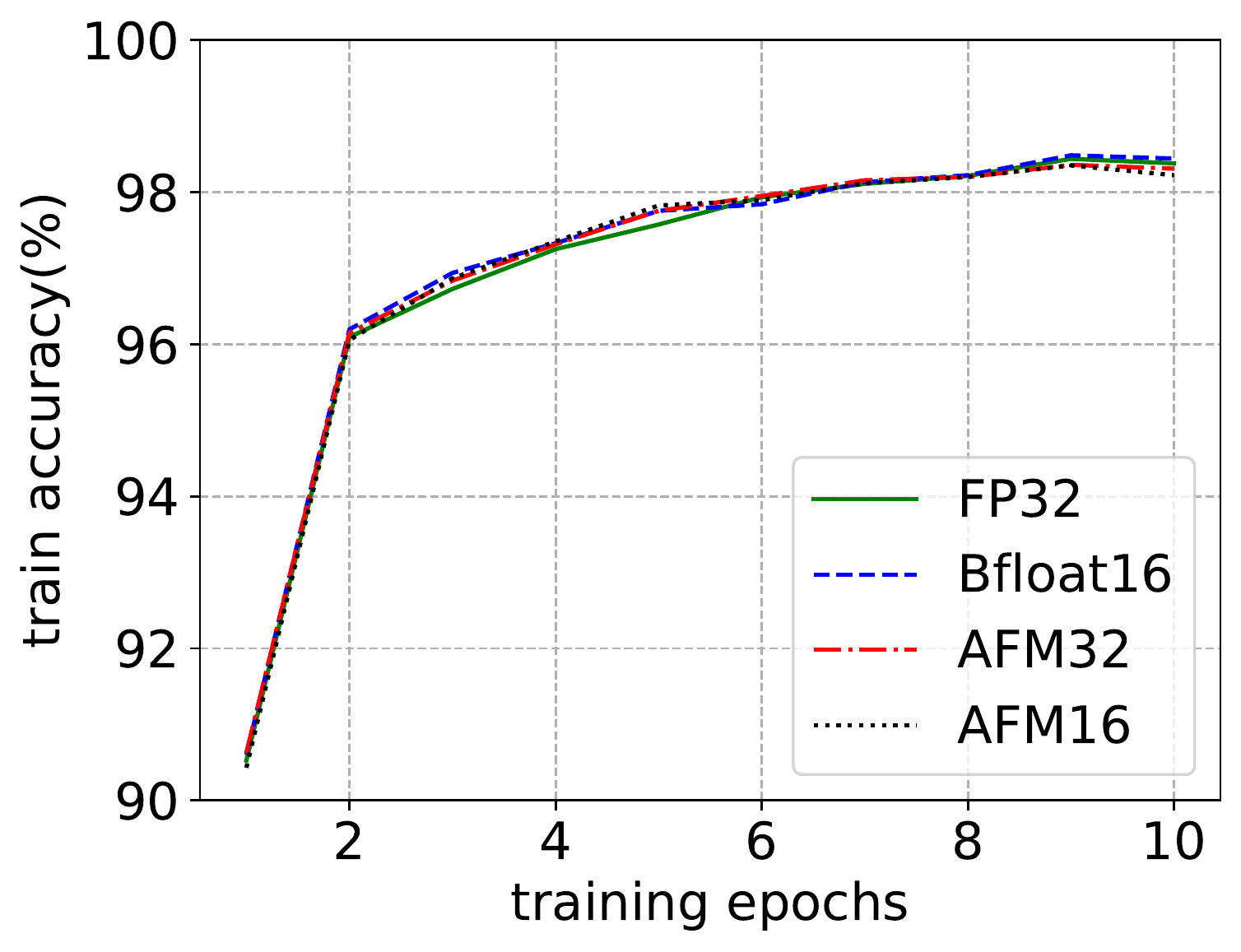}}  &  
\resizebox{0.29\textwidth}{!}{\includegraphics{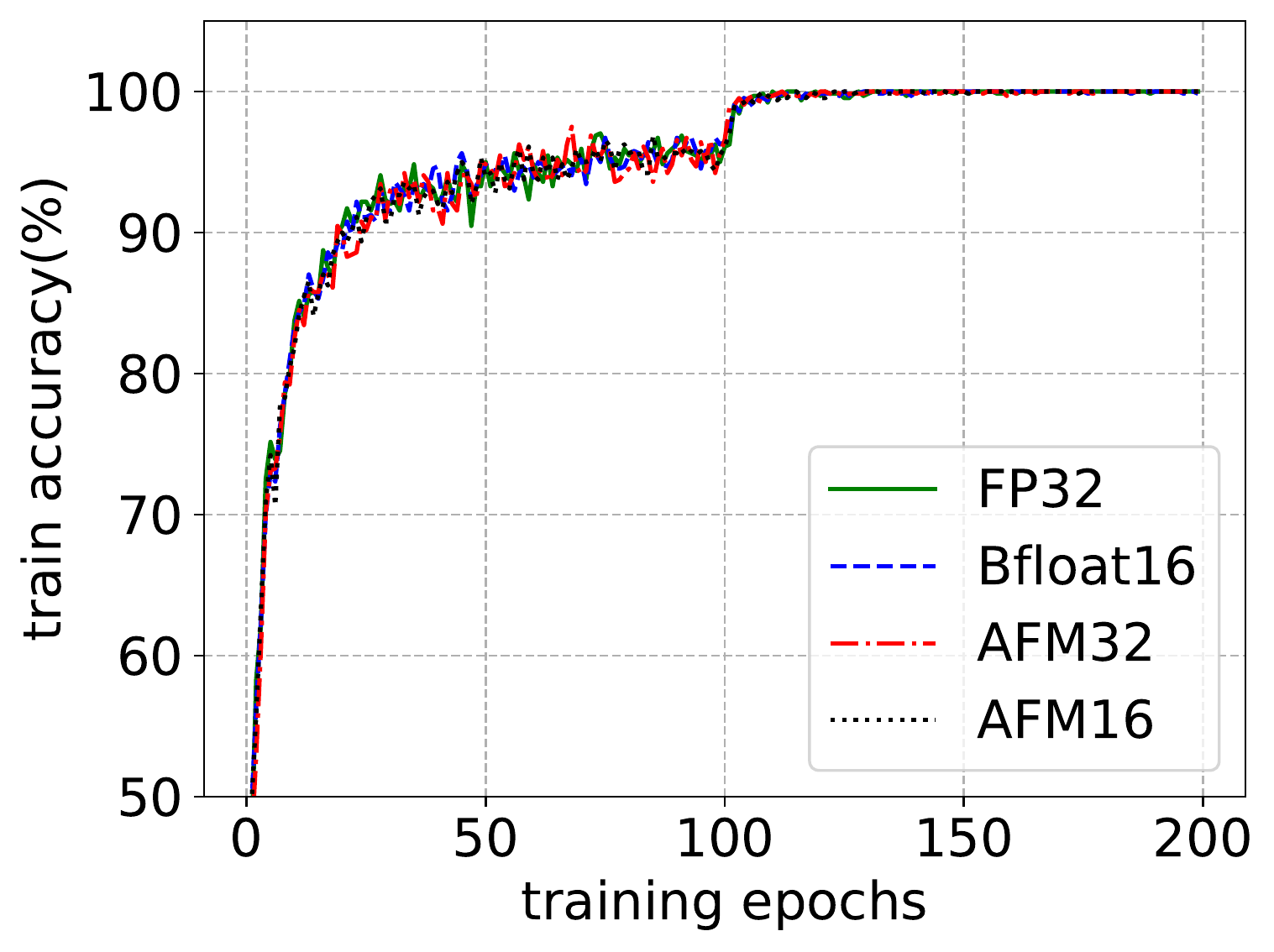}} \\
{ (d) CIFAR-10/ResNet34} &  { (e) CIFAR-10/ResNet50} & { (f) ImageNet/ResNet50 } \\
\resizebox{0.29\textwidth}{!}{\includegraphics{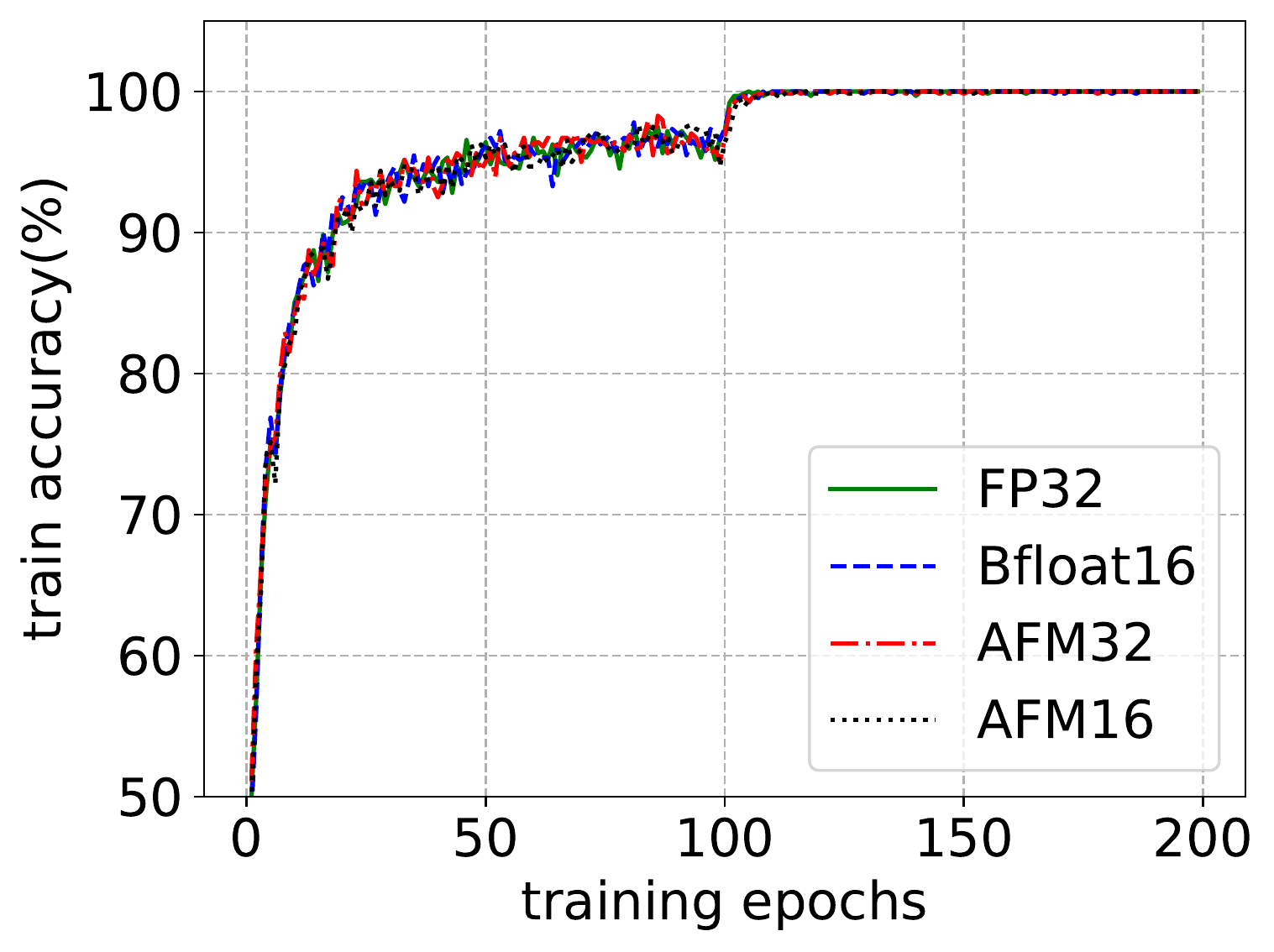}}  & \resizebox{0.29\textwidth}{!}{\includegraphics{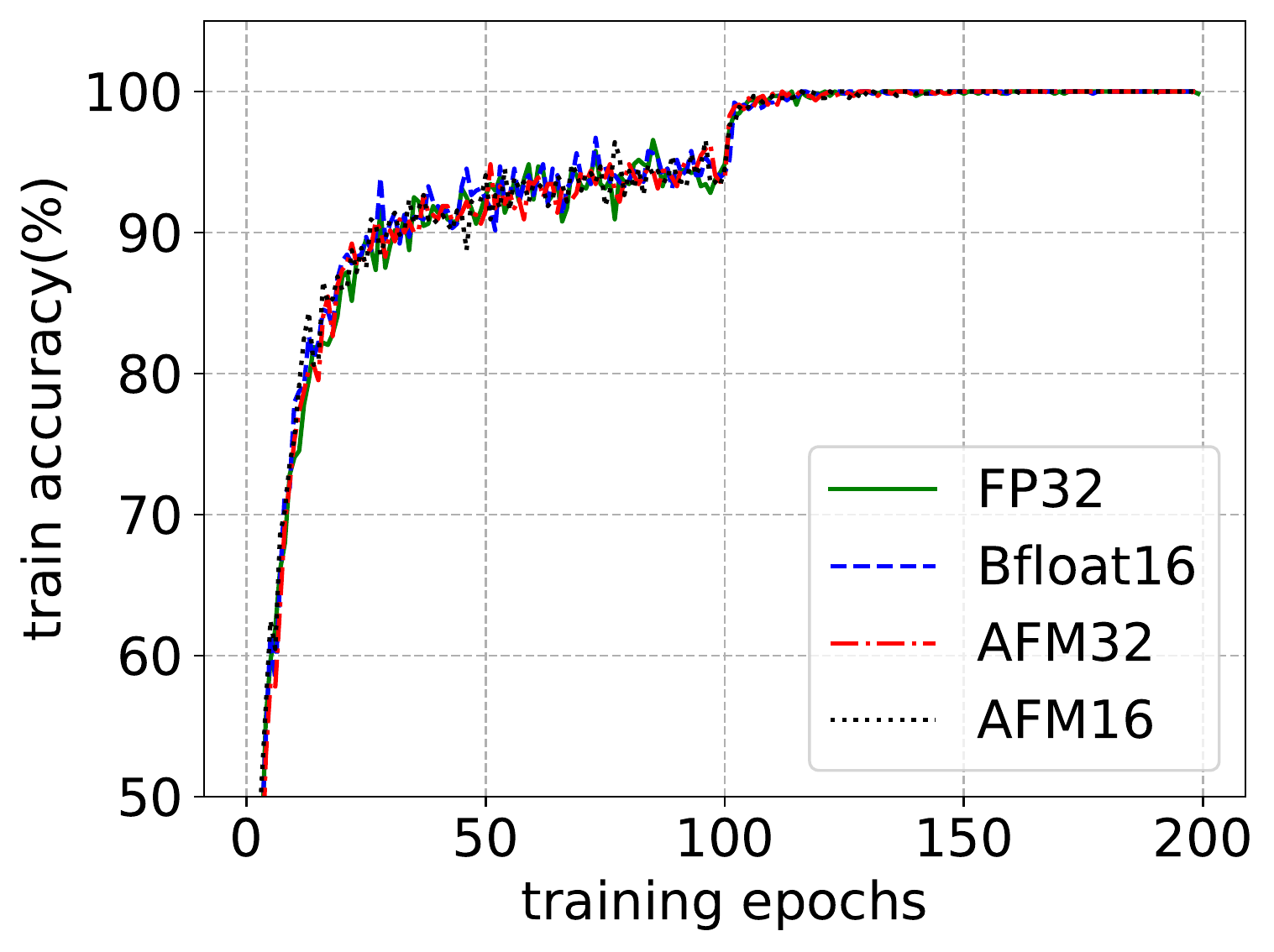}}  &  
\resizebox{0.29\textwidth}{!}{\includegraphics{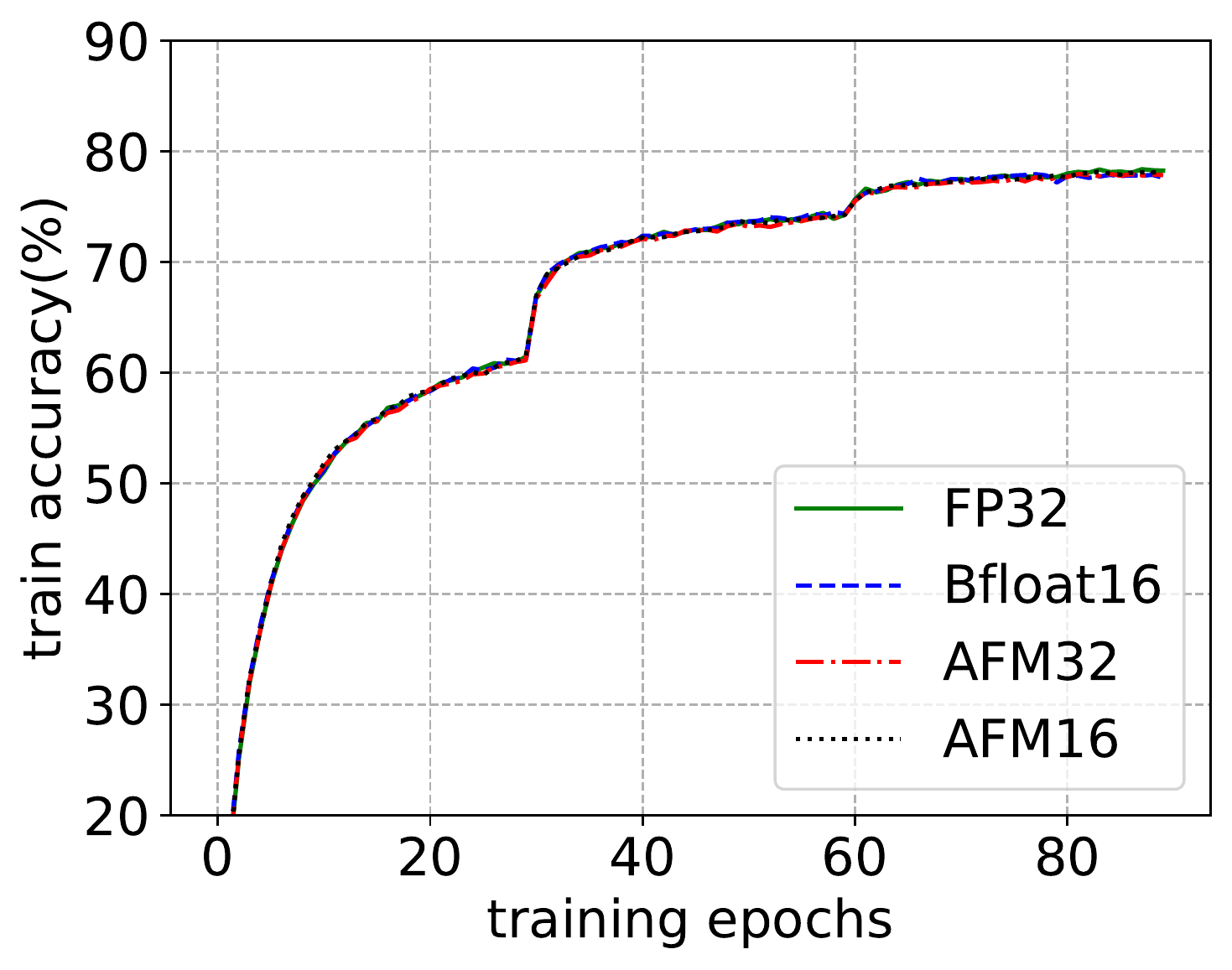}} \\
\end{tabular}
\caption{ \label{trainingcurve} Training curves for evaluated datasets and architectures using FP32, bfloat16 and approximate multipliers (AFM32 and AFM16). The convergence behaviour and convergence rate for AFM32 and AFM16 is similar to FP32 and bfloat16.}
\end{center}
\end{figure*}



\begin{table}[]
\caption{Data-types and multipliers used in experiments.}
\begin{footnotesize}
\begin{tabular}{|l|c| p{5.0cm}|}

\hline
\Tstrut
\textbf{Multiplier/} & \textbf{Bit-width} & \textbf{Description} \\
\textbf{Datatype} & \textbf{\textit{(s,e,m)}}  & \\
\hline
\hline

\Tstrut
FP32 &  (1,8,23)      & IEEE 754 standard format\\ \hline

\Tstrut
bfloat16 & (1,8,7)       & Brain Floating Point format~\cite{wang_bfloat16_2019} \\ \hline

\Tstrut
AFM32  & (1,8,23)  & 32-bit version of approx. mult AFM~\cite{saadat_minimally_2018}                \\ \hline

\Tstrut
AFM16 & (1,8,7) & 16-bit version of approx. mult AFM~\cite{saadat_minimally_2018}                                         \\ \hline

\end{tabular}

\end{footnotesize}
\label{tab_dtandmul}
\end{table}

\section{Results: Training Accuracy Evaluation}
\label{sec_results_accuracy}
 In this section, we present the evaluation of training accuracy and convergence using the approximate multipliers~\cite{saadat_minimally_2018} with ApproxTrain. 
For the following experiments, AFM32 and AFM16~\cite{saadat_minimally_2018} are used as representative approximate multipliers, whereas FP32 and bfloat16 formats are used as the baseline. \textcolor{black}{Figure \ref{fig2} depicted area and power efficiency of AFM16 and AFM32. In comparison with the FP32 multiplier, the AFM32 is 12x smaller and 24x more energy efficient, while the AFM16 is about 20x smaller and 50x more energy efficient.}
The different combinations of dataset/NN-architectures used in the experiments are listed as the title of each graph in Figure~\ref{trainingcurve} (and also listed in the first two columns of Table~\ref{tab:finalacc1}). For example, in Figure~\ref{trainingcurve} (a), MNIST dataset is used with LeNet-300-100 architecture.

\subsection{Training Convergence and Test Accuracy}

The training accuracy and convergence are depicted in Figure~\ref{trainingcurve}, where the training accuracy (y-axis) is plotted against train-epochs (x-axis) for the four multipliers listed in Table~\ref{tab_dtandmul}. The weights and parameters for the NN architectures are randomly initialized; however, for a given NN/dataset combination, the same random seed is used for all four multipliers (for fair comparison among different multipliers).
The training is run for several epochs until the validation accuracy stabilizes.
The training converges in 20 or fewer epochs for the two LeNets, whereas it stabilizes in around 100 epochs for CIFAR-10/ResNet combinations.
From Figure~\ref{trainingcurve}, we observe that the training-accuracy plots for the AFM32 and AFM16 closely follow the plots for FP32 and bfloat16. The observation applies to both the small datasets as well as the large dataset (ImageNet) training. In other words, training converges with approximate multipliers (AFM32 and AFM16), and the convergence behavior and convergence rate are the same as for FP32 and bfloat16. 
Note that, as shown in Figure~\ref{fig2}, AFM32 and AFM16 are much smaller and more power-efficient than FP32 and bfloat16 multipliers.

\begin{table}[!t]
\setlength\tabcolsep{2.8pt}
\caption{Test accuracy results for training with different multipliers. All Results are in Percentage(\%).}
\label{tab:finalacc1}
\begin{tabular}{|l l cccccc|}
\hline
                                   & \multicolumn{1}{l|}{}                                                                                     & \multicolumn{3}{c|}{\textbf{ \Tstrut 32-bit multipliers}}                                             & \multicolumn{3}{c|}{\textbf{16-bit multipliers}}                           \\ \cline{3-8} 
\multirow{-2}{*}{\textbf{Dataset}} & \multicolumn{1}{l|}{\multirow{-2}{*}{\textbf{\begin{tabular}[c]{@{}l@{}}Neural \\ Network\end{tabular}}}} & {FP32}  & {\Tstrut AFM32} & \multicolumn{1}{c|}{\textit{{diff}}}                & {bfloat16} & {AFM16} & \textit{{diff}}                \\ \hline
\multicolumn{8}{c}{\Tstrut Small Dataset}                                                                                                                                                                                                                                                                                        \\ \hline \hline
\Tstrut MNIST                              & \multicolumn{1}{l|}{LeNet-300-100}                                                                        & \textbf{96.90} & 97.10          & \multicolumn{1}{c|}{{\color[HTML]{3531FF} \textit{0.20}}}  & \textbf{96.70}    & 96.80          & {\color[HTML]{3531FF} \textit{0.10}}  \\
MNIST                              & \multicolumn{1}{l|}{LeNet-5}                                                                              & \textbf{98.30} & 98.30          & \multicolumn{1}{c|}{{\color[HTML]{3531FF} \textit{0.00}}}  & \textbf{98.30}    & 98.30          & {\color[HTML]{3531FF} \textit{0.00}}  \\
CIFAR10                            & \multicolumn{1}{l|}{ResNet18}                                                                             & \textbf{93.22} & 93.23          & \multicolumn{1}{c|}{{\color[HTML]{3531FF} \textit{0.01}}}  & \textbf{93.48}    & 93.40          & {\color[HTML]{680100} \textit{-0.08}} \\
CIFAR10                            & \multicolumn{1}{l|}{ResNet34}                                                                             & \textbf{93.51} & 93.57          & \multicolumn{1}{c|}{{\color[HTML]{3531FF} \textit{0.06}}}  & \textbf{93.73}    & 93.85          & {\color[HTML]{3531FF} \textit{0.12}}  \\
CIFAR10                            & \multicolumn{1}{l|}{ResNet50}                                                                             & \textbf{93.54} & 93.48          & \multicolumn{1}{c|}{{\color[HTML]{680100} \textit{-0.06}}} & \textbf{93.45}    & 93.62          & {\color[HTML]{3531FF} \textit{0.17}}  \\ \hline
\multicolumn{8}{c}{\Tstrut Large Dataset}                                                                                                                                                                                                                                                                                        \\ \hline \hline
\Tstrut ImageNet                           & \multicolumn{1}{l|}{ResNet50}                                                                             & \textbf{73.10} & 73.00          & \multicolumn{1}{c|}{{\color[HTML]{680100} \textit{-0.10}}} & \textbf{73.10}    & 73.10          & {\color[HTML]{3531FF} \textit{0.00}}  \\ \hline
\end{tabular}
\end{table} 

The final test accuracy results for the six dataset/architecture combinations are reported in Table~\ref{tab:finalacc1}, for 32-bit and 16-bit formats. The third and sixth columns, presenting results for FP32 and bfloat16, are considered are baselines for 32-bit and 16-bit formats, respectively. The difference of test accuracy between AFMs compared to the corresponding baselines is listed in the fifth and eighth columns.
From Table~\ref{tab:finalacc1}, for both data formats, we observe that the test accuracy for all dataset/architecture combinations using approximate multipliers is very similar to the baseline  (accuracy degradation is within 0.10\%). Note that such accuracy differences also exist between the heavily adapted FP32 and bfloat16 formats (column 3 and 6--Table~\ref{tab:finalacc1}). Therefore, we argue that such degradation is acceptable.
In fact, in most cases, the accuracy for approximate multipliers is slightly better than the baselines (highlighted in blue in the table). A reason for this is that the error injected by erroneous approximate multiplications (AFMs) in training can be considered as stochastic noise, which is a type of regularization \cite{noh_regularizing_2017}.

\subsection{Cross-format Test Accuracy}
\begin{table}[!th]
\centering
\setlength\tabcolsep{2.0pt}
\caption{Cross format Testing for Resnet50-ImageNet. All results are in percentage(\%).}
\label{tab:crossformat}
\begin{tabular}{|rc||cccc|}
\hline
\multicolumn{2}{|c||}{}                                                                                               & \multicolumn{4}{c|}{\textit{\textbf{\Tstrut used for testing}}}                                                                   \\
\multicolumn{2}{|c||}{\multirow{-2}{*}{\textit{\textbf{Multipliers}}}}                                                & \textbf{\Tstrut  FP32}                & \textbf{AFM32}               & \textbf{bfloat16}            & \textbf{AFM16}               \\ \hhline{|==#====|}
                                                                                                 & \textbf{\Tstrut  FP32}     & \textbf{73.10}               & {\color[HTML]{3531FF} 73.10} & {\color[HTML]{3531FF} 73.10} & {\color[HTML]{9A0000} 73.00} \\
                                                                                                 & \textbf{\Tstrut AFM32}    & {\color[HTML]{3531FF} 73.00} & \textbf{73.00}               & {\color[HTML]{3531FF} 73.00} & {\color[HTML]{3531FF} 73.10} \\
                                                                                                 & \textbf{\Tstrut bfloat16} & {\color[HTML]{3531FF} 73.00} & {\color[HTML]{3531FF} 73.00} & \textbf{73.10}               & {\color[HTML]{3531FF} 73.00} \\
\multirow{-4}{*}{\textit{\textbf{\begin{tabular}[c]{@{}r@{}}used for \\ training\end{tabular}}}} & \textbf{\Tstrut AFM16}    & {\color[HTML]{9A0000} 73.00} & {\color[HTML]{9A0000} 73.00} & {\color[HTML]{9A0000} 73.00} & \textbf{73.10}               \\ \hline
\end{tabular}
\end{table}

\begin{figure}[th]
    \centering
    \includegraphics[width=\linewidth]{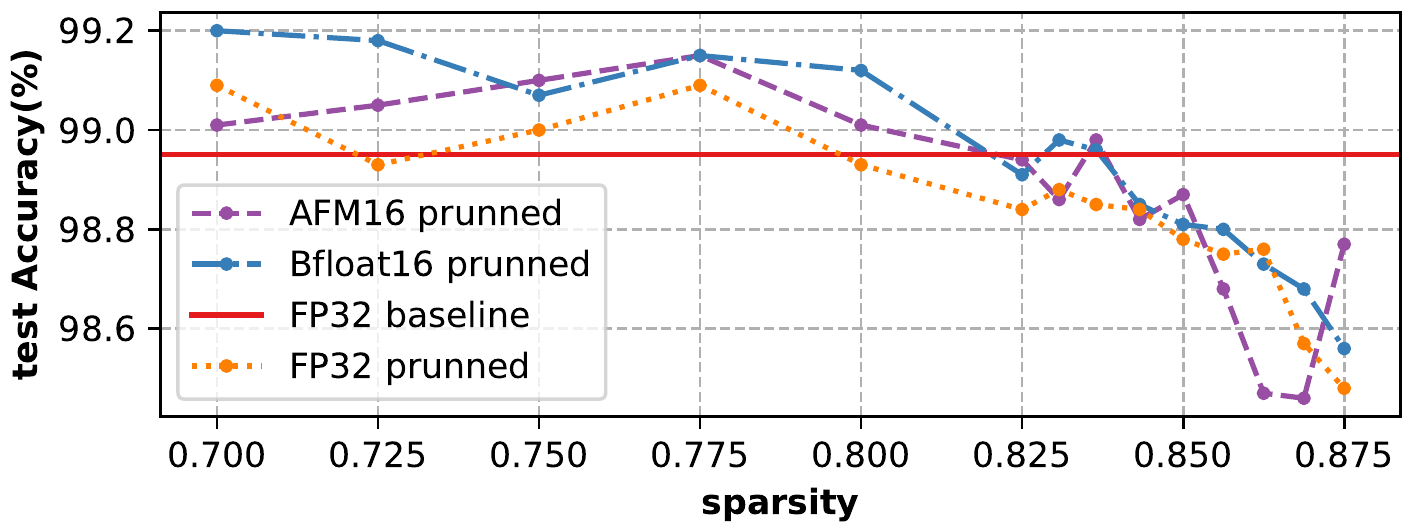}
    \caption{MNIST CNN pruning results with different multipliers.}
    \label{fig_TresultsLenet}
\end{figure}
For the ImageNet dataset, we perform another experiment where we evaluate the test accuracy with a multiplier that is different from the one used for training. In other words, we train the neural network using one multiplier type and test it using another. The purpose of this experiment is to observe if any drastic over-fitting occurs w.r.t. the used multiplier type.

The results of the experiment are listed in Table~\ref{tab:crossformat}. The multipliers used for training are listed along the second column, while the multipliers listed across the second row are used for testing. Essentially, the numbers in the diagonal (highlighted in bold) are test results when the same multiplier is used for training and test and are the baseline for each row. The rest of the entries in Table~\ref{tab:crossformat} depict test-accuracy results when different multipliers are used for testing. We observe that the difference in accuracy is within 0.10\%, which we deem acceptable, as discussed in the previous subsection.
Therefore, this experiment demonstrates that we may safely train and deploy a neural network with different multiplier types (including approximate multipliers) as they do not drastically over-fit for the given data/multiplier type.

\subsection{Approximate Multiplier on top of Pruning}

We also performed an experiment to couple pruning with use of approximate multipliers in training. Pruning is a mechanism for efficient inference, and involved repeated training effort. Thus, it is beneficial to improve the training efficiency and demonstrate that our framework enables hardware/algorithm co-design. The pruning code/algorithm is implemented following the official TensorFlow example without any custom modifications. The pruning schedule is polynomial decay. The initial sparsity is set to 70\% and final sparsity is set to different levels to find the optimal sparsity. First, a CNN with 2 convolution layers and three dense layers was pre-trained for 20 epochs. Then, the weights of the pre-trained CNN is loaded into a new model that to be pruned. After every pruning, the model is retrained for another two epochs to refine accuracy.
In Figure \ref{fig_TresultsLenet}, the red horizontal dash-dot line represents baseline for all the other experiments. The orange, blue and purple curves are pruned test accuracy against sparsity for FP32, bfloat16 and AFM16, respectively.
Overall, these curves slowly declined from 70\% to 80\% sparsity and dropped rapidly after 80\% sparsity. We observed all three curves are above the baseline from 70\% to 80\%. This caused by sparse weights acting as dropout layer, providing extra regularization to help the model generalize. It can be observed that 83\% sparsity level is optimal for pruned bfloat16 and pruned AFM16 as they are higher or equal to the baseline. However, FP32 dropped below the baseline with 83\% sparsity. Additionally, curves of AFM16 and bfloat16 are consistently above the baseline, demonstrating that AFM16 could act as a drop-in replacement for native bfloat16 multiplier.
In this experiment, we successfully coupled approximate multiplier designs with pruning algorithm, thus highlighting the flexibility of our framework.

\begin{table*}[!t]
\caption{Training run-time results on System-I and System-II.}
\scriptsize
\setlength\tabcolsep{3.0pt}
\begin{tabular}{|ll||rrrrccc||rrrrccc|}
\hline
                                   &                                                                                      & \multicolumn{7}{c||}{\Tstrut\textbf{System - I (V100 GPU) }}                                                                                                                                                                                                                                                                                                                                                                                                                                                                                                                          & \multicolumn{7}{c|}{\textbf{System - II (GTX1080 GPU) }}                                                                                                                                                                                                                                                                                                                                                                                                                                                                                                                         \\ \cline{3-16} 
                                   &                                                                                      & \multicolumn{4}{c|}{\Tstrut\textbf{ Actual Time per batch}}                                                                                                                                                                                                       & \multicolumn{3}{c||}{\textbf{Speed Ratio}}                                                                                                                                                                                                                                                           & \multicolumn{4}{c|}{\textbf{Actual Time per batch}}                                                                                                                                                                                                       & \multicolumn{3}{c|}{\textbf{Speed Ratio}}                                                                                                                                                                                                                                                           \\ \cline{3-16} 
                                   &                                                                                      & \multicolumn{1}{c|}{\begin{tabular}[c]{@{}c@{}}\Tstrut TF with \\ native mult.\end{tabular}} & \multicolumn{1}{c|}{\begin{tabular}[c]{@{}c@{}}AT with \\ native mult.\end{tabular}} & \multicolumn{2}{c|}{\begin{tabular}[c]{@{}c@{}}AT with \\ AFM\end{tabular}} & \multicolumn{1}{c|}{}                                                                                  & \multicolumn{1}{c|}{}                                                                                  &                                                                                   & \multicolumn{1}{c|}{\begin{tabular}[c]{@{}c@{}}TF with \\ native mult.\end{tabular}} & \multicolumn{1}{c|}{\begin{tabular}[c]{@{}c@{}}AT with \\ native mult.\end{tabular}} & \multicolumn{2}{c|}{\begin{tabular}[c]{@{}c@{}}AT with \\ AFM\end{tabular}} & \multicolumn{1}{c|}{}                                                                                  & \multicolumn{1}{c|}{}                                                                                  &                                                                                   \\ \cline{5-6} \cline{12-13}
                                   &                                                                                      & \multicolumn{1}{c|}{GPU}                                                             & \multicolumn{1}{c|}{\Tstrut GPU}                                                             & \multicolumn{1}{c}{GPU}              & \multicolumn{1}{c|}{CPU}             & \multicolumn{1}{c|}{}                                                                                  & \multicolumn{1}{c|}{}                                                                                  &                                                                                   & \multicolumn{1}{c|}{GPU}                                                             & \multicolumn{1}{c|}{GPU}                                                             & \multicolumn{1}{c}{GPU}              & \multicolumn{1}{c|}{CPU}             & \multicolumn{1}{c|}{}                                                                                  & \multicolumn{1}{c|}{}                                                                                  &                                                                                   \\
\multirow{-5}{*}{\textbf{DataSet}} & \multirow{-5}{*}{\textbf{\begin{tabular}[c]{@{}l@{}}Neural \\ Network\end{tabular}}} & \multicolumn{1}{c|}{(TFnG)}                                                          & \multicolumn{1}{c|}{(ATnG)}                                                          & \multicolumn{1}{c}{(ATxG)}           & \multicolumn{1}{c|}{(ATxC)}          & \multicolumn{1}{c|}{\multirow{-3}{*}{\begin{tabular}[c]{@{}c@{}}ATnG/\\ TFnG\\ (slower)\end{tabular}}} & \multicolumn{1}{c|}{\multirow{-3}{*}{\begin{tabular}[c]{@{}c@{}}ATxG/\\ TFnG\\ (slower)\end{tabular}}} & \multirow{-3}{*}{\begin{tabular}[c]{@{}c@{}}ATxC/\\ ATxG\\ (faster)\end{tabular}} & \multicolumn{1}{c|}{(TFnG)}                                                          & \multicolumn{1}{c|}{(ATnG)}                                                          & \multicolumn{1}{c}{(ATxG)}           & \multicolumn{1}{c|}{(ATxC)}          & \multicolumn{1}{c|}{\multirow{-3}{*}{\begin{tabular}[c]{@{}c@{}}ATnG/\\ TFnG\\ (slower)\end{tabular}}} & \multicolumn{1}{c|}{\multirow{-3}{*}{\begin{tabular}[c]{@{}c@{}}ATxG/\\ TFnG\\ (slower)\end{tabular}}} & \multirow{-3}{*}{\begin{tabular}[c]{@{}c@{}}ATxC/\\ ATxG\\ (faster)\end{tabular}} \\ 

\hline 
\hline

\Tstrut
MNIST                              & {\tiny LeNet-300-100}                                                                & \textit{2.0 ms}                                                                      & 3 ms                                                                               & 3 ms                               & \multicolumn{1}{r|}{3 s}             		& \textbf{1.3${\times}$}                                                                                         & {\color[HTML]{3531FF} \textbf{1.6${\times}$}}                                                                  & {\color[HTML]{036400} \textbf{884${\times}$}}                                            & \textit{0.9 ms}                                                                    & 2 ms                                                                                 & 3 ms                                         & \multicolumn{1}{r|}{3 s}             & \textbf{2.3${\times}$}                                                                                         & {\color[HTML]{3531FF} \textbf{3.2${\times}$}}                                                                  & {\color[HTML]{036400} \textbf{1078${\times}$} }                                                                  \\
MNIST                              & LeNet-5                                                                              & \textit{3 ms}                                                                        & 7 ms                                                                                 & 13 ms                                & \multicolumn{1}{r|}{23 s}            & \textbf{2.3${\times}$}                                                                                         & {\color[HTML]{3531FF} \textbf{4.2${\times}$}}                                                                  & {\color[HTML]{036400} \textbf{1798${\times}$}}                                            & \textit{2 ms}                                                                        & 9 ms                                                                                 & 16 ms                                   & \multicolumn{1}{r|}{22 s}            & \textbf{4.4${\times}$}                                                                                         & {\color[HTML]{3531FF} \textbf{8.3${\times}$}}                                                                  & {\color[HTML]{036400} \textbf{1374${\times}$} }                                                                  \\
CIFAR10                            & ResNet18                                                                             & \textit{13 ms}                                                                       & 49 ms                                                                                & 178 ms                               & \multicolumn{1}{r|}{736 s}           & \textbf{3.7${\times}$}                                                                                         & {\color[HTML]{3531FF} \textbf{13.5${\times}$}}                                                                 & {\color[HTML]{036400} \textbf{4132${\times}$}}                                            & \textit{27 ms}                                                                       & 120 ms                                                                                & 357 ms                                 & \multicolumn{1}{r|}{672 s}           & \textbf{4.4${\times}$}                                                                                         & {\color[HTML]{3531FF} \textbf{13.1${\times}$}}                                                                 & {\color[HTML]{036400} \textbf{1882${\times}$}}                                                                    \\
CIFAR10                            & ResNet34                                                                             & \textit{23 ms}                                                                       & 90 ms                                                                                & 338 ms                               & \multicolumn{1}{r|}{1376 s}          & \textbf{4.0${\times}$}                                                                                         & {\color[HTML]{3531FF} \textbf{15.0${\times}$}}                                                                 & {\color[HTML]{036400} \textbf{4072${\times}$}}                                            & \textit{49 ms}                                                                       & 221 ms                                                                               & 682 ms                                  & \multicolumn{1}{r|}{1280 s}          & \textbf{4.5${\times}$}                                                                                         & {\color[HTML]{3531FF} \textbf{13.8${\times}$}}                                                                 & {\color[HTML]{036400} \textbf{1877${\times}$}}                                                                   \\
CIFAR10                            & ResNet50                                                                             & \textit{44 ms}                                                                       & 154 ms                                                                               & 478 ms                               & \multicolumn{1}{r|}{1632 s}          & \textbf{3.5${\times}$}                                                                                         & {\color[HTML]{3531FF} \textbf{10.8${\times}$}}                                                                 & {\color[HTML]{036400} \textbf{3417${\times}$}}                                            & \textit{93 ms}                                                                       & 366 ms                                                                               & 960 ms                                  & \multicolumn{1}{r|}{1376 s}          & \textbf{3.9${\times}$}                                                                                         & {\color[HTML]{3531FF} \textbf{10.3${\times}$}}                                                                 & {\color[HTML]{036400} \textbf{1433${\times}$} }                                                                   \\
ImageNet                           & ResNet50                                                                             & \textit{114 ms}                                                                      & 460 ms                                                                               & 1464ms                                  & \multicolumn{1}{r|}{4896 s}       & \textbf{4.0${\times}$}                                                                                         & {\color[HTML]{3531FF} \textbf{12.8${\times}$}}                                                                 & {\color[HTML]{036400} \textbf{3343${\times}$}}                                            & \textit{267 ms}                                                                      & 1186 ms                                                                               & 3091 s                                  & \multicolumn{1}{r|}{4864 s}          & \textbf{4.3${\times}$}                                                                                         & {\color[HTML]{3531FF} \textbf{11.6${\times}$}}                                                                 &{\color[HTML]{036400}  \textbf{1574${\times}$}    }                                                                \\ \hline
\end{tabular}
\label{tab_timingtraining}
\end{table*}

\begin{table*}[!t]
\caption{Inference  run-time results on System-I and System-II.}
\scriptsize
\setlength\tabcolsep{3.0pt}
\begin{tabular}{|ll||rrrrccc||rrrrccc|}
\hline
                                   &                                                                                      & \multicolumn{7}{c||}{\Tstrut\textbf{System - I (V100 GPU) }}                                                                                                                                                                                                                                                                                                                                                                                                                                                                                                                          & \multicolumn{7}{c|}{\textbf{System - II (GTX1080 GPU) }}                                                                                                                                                                                                                                                                                                                                                                                                                                                                                                                         \\ \cline{3-16} 
                                   &                                                                                      & \multicolumn{4}{c|}{\Tstrut\textbf{ Actual Time per batch}}                                                                                                                                                                                                       & \multicolumn{3}{c||}{\textbf{Speed Ratio}}                                                                                                                                                                                                                                                           & \multicolumn{4}{c|}{\textbf{Actual Time per batch}}                                                                                                                                                                                                       & \multicolumn{3}{c|}{\textbf{Speed Ratio}}                                                                                                                                                                                                                                                           \\ \cline{3-16} 
                                   &                                                                                      & \multicolumn{1}{c|}{\begin{tabular}[c]{@{}c@{}}\Tstrut TF with \\ native mult.\end{tabular}} & \multicolumn{1}{c|}{\begin{tabular}[c]{@{}c@{}}AT with \\ native mult.\end{tabular}} & \multicolumn{2}{c|}{\begin{tabular}[c]{@{}c@{}}AT with \\ AFM\end{tabular}} & \multicolumn{1}{c|}{}                                                                                  & \multicolumn{1}{c|}{}                                                                                  &                                                                                   & \multicolumn{1}{c|}{\begin{tabular}[c]{@{}c@{}}TF with \\ native mult.\end{tabular}} & \multicolumn{1}{c|}{\begin{tabular}[c]{@{}c@{}}AT with \\ native mult.\end{tabular}} & \multicolumn{2}{c|}{\begin{tabular}[c]{@{}c@{}}AT with \\ AFM\end{tabular}} & \multicolumn{1}{c|}{}                                                                                  & \multicolumn{1}{c|}{}                                                                                  &                                                                                   \\ \cline{5-6} \cline{12-13}
                                   &                                                                                      & \multicolumn{1}{c|}{GPU}                                                             & \multicolumn{1}{c|}{\Tstrut GPU}                                                             & \multicolumn{1}{c}{GPU}              & \multicolumn{1}{c|}{CPU}             & \multicolumn{1}{c|}{}                                                                                  & \multicolumn{1}{c|}{}                                                                                  &                                                                                   & \multicolumn{1}{c|}{GPU}                                                             & \multicolumn{1}{c|}{GPU}                                                             & \multicolumn{1}{c}{GPU}              & \multicolumn{1}{c|}{CPU}             & \multicolumn{1}{c|}{}                                                                                  & \multicolumn{1}{c|}{}                                                                                  &                                                                                   \\
\multirow{-5}{*}{\textbf{DataSet}} & \multirow{-5}{*}{\textbf{\begin{tabular}[c]{@{}l@{}}Neural \\ Network\end{tabular}}} & \multicolumn{1}{c|}{(TFnG)}                                                          & \multicolumn{1}{c|}{(ATnG)}                                                          & \multicolumn{1}{c}{(ATxG)}           & \multicolumn{1}{c|}{(ATxC)}          & \multicolumn{1}{c|}{\multirow{-3}{*}{\begin{tabular}[c]{@{}c@{}}ATnG/\\ TFnG\\ (slower)\end{tabular}}} & \multicolumn{1}{c|}{\multirow{-3}{*}{\begin{tabular}[c]{@{}c@{}}ATxG/\\ TFnG\\ (slower)\end{tabular}}} & \multirow{-3}{*}{\begin{tabular}[c]{@{}c@{}}ATxC/\\ ATxG\\ (faster)\end{tabular}} & \multicolumn{1}{c|}{(TFnG)}                                                          & \multicolumn{1}{c|}{(ATnG)}                                                          & \multicolumn{1}{c}{(ATxG)}           & \multicolumn{1}{c|}{(ATxC)}          & \multicolumn{1}{c|}{\multirow{-3}{*}{\begin{tabular}[c]{@{}c@{}}ATnG/\\ TFnG\\ (slower)\end{tabular}}} & \multicolumn{1}{c|}{\multirow{-3}{*}{\begin{tabular}[c]{@{}c@{}}ATxG/\\ TFnG\\ (slower)\end{tabular}}} & \multirow{-3}{*}{\begin{tabular}[c]{@{}c@{}}ATxC/\\ ATxG\\ (faster)\end{tabular}} \\

\hline 
\hline

\Tstrut MNIST                      & {\tiny LeNet-300-100}                                                                & \textit{1 ms}                                                                        & 1 ms                                                                                 & 2 ms                                 & \multicolumn{1}{r|}{1 s}             & \textbf{1.2${\times}$}                                                                                         & {\color[HTML]{3531FF} \textbf{1.5${\times}$}}                                                                  & {\color[HTML]{036400} \textbf{609${\times}$}}                                             & \textit{0.869 ms}                                                                    & 1 ms                                                                             & 1 ms                                 & \multicolumn{1}{r|}{857 ms}          & \textbf{1.3${\times}$}                                                                                         & {\color[HTML]{3531FF} \textbf{1.4${\times}$}}                                                                  &{\color[HTML]{036400}  \textbf{697${\times}$}}                                                                    \\
MNIST                              & LeNet-5                                                                              & \textit{2 ms}                                                                        & 3 ms                                                                                 & 4 ms                                 & \multicolumn{1}{r|}{8 s}              & \textbf{1.7${\times}$}                                                                                         & {\color[HTML]{3531FF} \textbf{2.1${\times}$}}                                                                  & {\color[HTML]{036400} \textbf{1780${\times}$}}                                            & \textit{1 ms}                                                                        & 3 ms                                                                                 & 4 ms                                 & \multicolumn{1}{r|}{7 s}             & \textbf{2.5${\times}$}                                                                                         & {\color[HTML]{3531FF} \textbf{3.6${\times}$}}                                                                  & {\color[HTML]{036400} \textbf{1815${\times}$} }                                                                  \\
CIFAR10                            & ResNet18                                                                             & \textit{5 ms}                                                                        & 15 ms                                                                                & 56 ms                                & \multicolumn{1}{r|}{320 s}           & \textbf{3.0${\times}$}                                                                                         & {\color[HTML]{3531FF} \textbf{11.3${\times}$}}                                                                 & {\color[HTML]{036400} \textbf{5743${\times}$}}                                            & \textit{8 ms}                                                                        & 39 ms                                                                                & 113 ms                               & \multicolumn{1}{r|}{352 s}           & \textbf{4.8${\times}$}                                                                                         & {\color[HTML]{3531FF} \textbf{13.7${\times}$}}                                                                 & {\color[HTML]{036400} \textbf{3102${\times}$} }                                                                  \\
CIFAR10                            & ResNet34                                                                             & \textit{9 ms}                                                                        & 25 ms                                                                                & 107 ms                               & \multicolumn{1}{r|}{576 s}           & \textbf{2.9${\times}$}                                                                                         & {\color[HTML]{3531FF} \textbf{12.2${\times}$}}                                                                 & {\color[HTML]{036400} \textbf{5405${\times}$}}                                            & \textit{15 ms}                                                                       & 68 ms                                                                                & 217 ms                               & \multicolumn{1}{r|}{640 s}           & \textbf{4.5${\times}$}                                                                                         & {\color[HTML]{3531FF} \textbf{14.3${\times}$}}                                                                 &{\color[HTML]{036400}  \textbf{2952${\times}$}}                                                                   \\
CIFAR10                            & ResNet50                                                                             & \textit{14 ms}                                                                       & 36 ms                                                                                & 131 ms                               & \multicolumn{1}{r|}{544 s}           & \textbf{2.5${\times}$}                                                                                         & {\color[HTML]{3531FF} \textbf{9.2${\times}$}}                                                                 & {\color[HTML]{036400} \textbf{4154${\times}$}}                                            & \textit{26 ms}                                                                       & 93 ms                                                                                & 265 ms                               & \multicolumn{1}{r|}{512 s}           & \textbf{3.6${\times}$}                                                                                         & {\color[HTML]{3531FF} \textbf{10.1${\times}$}}                                                                 & {\color[HTML]{036400} \textbf{1934${\times}$}}                                                                   \\
ImageNet                           & ResNet50                                                                             & \textit{35 ms}                                                                       & 110 ms                                                                               & 398 ms                               & \multicolumn{1}{r|}{1580 s}          & \textbf{3.2${\times}$}                                                                                         & {\color[HTML]{3531FF} \textbf{11.4${\times}$}}                                                                 & {\color[HTML]{036400} \textbf{3993${\times}$}}                                            & \textit{74 ms}                                                                       & 301 ms                                                                               & 855 ms                                   & \multicolumn{1}{r|}{1568 s}          & \textbf{4${\times}$}                                                                                         & {\color[HTML]{3531FF} \textbf{11.5${\times}$}}                                                                 & {\color[HTML]{036400} \textbf{1833${\times}$} }                                                                   \\ \hline
\end{tabular}
\label{tab_timinginference}
\end{table*}
\section{Results: Runtime Performance Evaluation}
\label{sec_results_performance}
As discussed in Section I,  the aim of ApproxTrain is to perform DNN training with approximate multiplier simulation in practically feasible run-times.
In this section, we present results for a detailed evaluation of the timing performance of ApproxTrain for \textit{training} as well as for \textit{inference}.

The overall timing performance comparison is evaluated by recording the average time for DNNs to train/infer one batch.
The results are listed in Table~\ref{tab_timingtraining} and Table~\ref{tab_timinginference}.
For these evaluations, the training and inference experiments are run on two platforms: System-I and System-II (described in Section~\ref{sec_expsetup}).
For both training  and inference, and for both platforms, Table~\ref{tab_timingtraining}\&\ref{tab_timinginference} present four types of run-time measurements. These are: 

\smallskip

\begin{enumerate}[noitemsep,nolistsep,leftmargin=*]
\item \textit{TFnG} -- run-time for training/inference performed using standard TensorFlow  with cuDNN/cuBLAS libraries on GPU with native hardware multiplier (FP32); 
\item \textit{ATnG} -- run-time for training/inference performed using ApproxTrain with custom CUDA kernels (described in Section~\ref{sec_customcuda}) on GPU with native hardware multiplier (FP32);
\item \textit{ATxG} -- run-time for training/inference performed using ApproxTrain with custom CUDA kernels on GPU with AMSim (16-bit FP datatype (1, 8, 7) in Table \ref{tab_dtandmul}); and,
\item \textit{ATxC} -- run-time for training/inference performed using ApproxTrain with custom CUDA kernels on CPU \textcolor{black}{with direct C/C++ simulation of approx. multiplier.}
\end{enumerate}

\smallskip

The \textit{TFnG} values, i.e., the run-times of standard TensorFlow  with native hardware multiplication supported by GPU, are considered as the baseline in the following discussion.
\textit{Note that we did not perform run-time evaluation experiments with bfloat16 since the available hardware did not natively support bfloat16. Meanwhile, 16-bit FP datatype (1, 8, 7) as shown in Table \ref{tab_dtandmul} is used in AMSim, considering it equivalent to the industry de-facto standard for training/inference.}

\smallskip

\subsubsection{\textbf{Custom CUDA kernels in ApproxTrain vs optimized cuDNN/cuBLAS in TensorFlow}}
For this comparison, we use ApproxTrain with the `*' operator for multiplication, which invokes the native hardware multiplier on GPU instead of an approximate multiplier simulation model. \textcolor{black}{This comparison demonstrates the performance of custom kernels on GPUs with native multiplier hardware.} Therefore, in columns 4 \& 11 of Table~\ref{tab_timingtraining}\&\ref{tab_timinginference}, \textit{ATnG} refers to the time with our custom CUDA kernels in ApproxTrain, as opposed to columns 3 \& 10 which refers to the time taken by cuDNN/cuBLAS based TensorFlow.  

The slow-down (speed-ratio) of \textit{ATnG} compared with \textit{TFnG} is highlighted in bold (black) in Table~\ref{tab_timingtraining}\&\ref{tab_timinginference}.
In the training phase, ApproxTrain with native multiplication is
$1{\times}$ -- $5{\times}$ slower than standard TensorFlow for the various datasets. Note that the closed-source cuDNN and cuBLAS libraries have been optimized by teams of several hundred professionals within Nvidia for over a decade. Thus, we believe that less than $5{\times}$ slow-down is reasonable. Nonetheless, since our framework is open-source, the research community may contribute with further optimizations.
\begin{figure}[t]
    \centering
    \includegraphics[width=\linewidth]{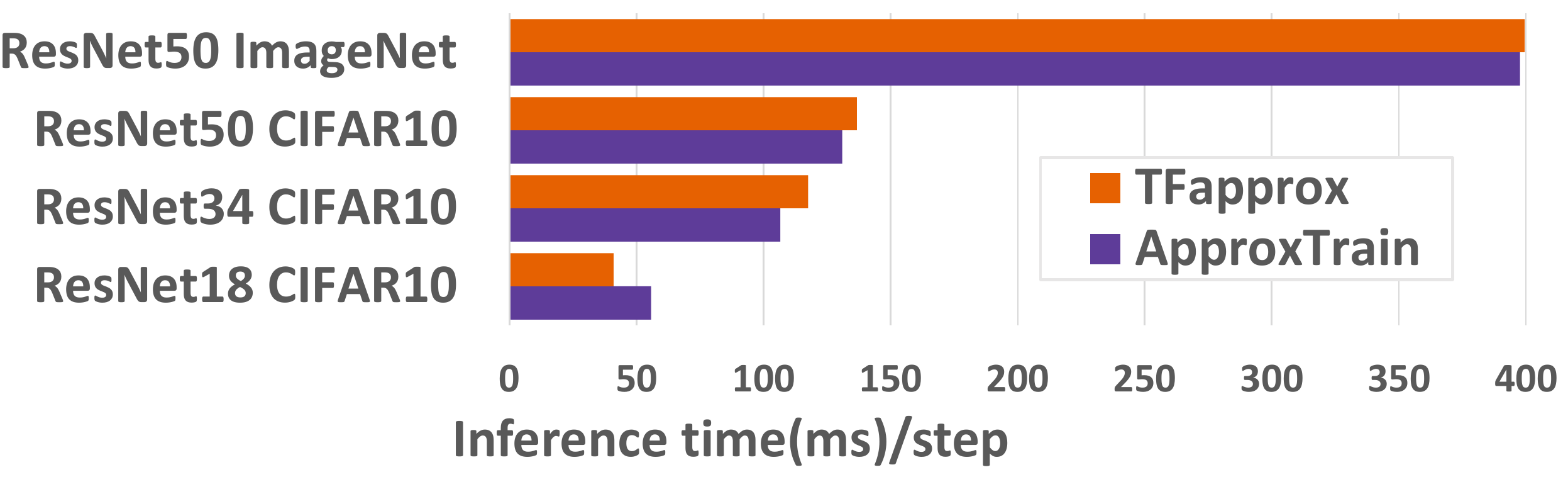}
    \caption{Inference performance comparison for ApproxTrain and TFapprox.}
    \label{fig:run_time_perf_cmp}
\end{figure}

\smallskip

\subsubsection{\textbf{ApproxTrain perf. with approx. multiplier simulation}}

We compare the run-times of ApproxTrain with approximate multiplier simulation on GPU \textit{(ATxG)} against the standard TF with native multiplication on GPU (\textit{TFnG}). The slow-down (speed-ratio) of this comparison is highlighted in bold (blue) in Table~\ref{tab_timingtraining}\&\ref{tab_timinginference}.
Slow-down is around $2{\times}$ for the smallest data-set/architecture, whereas the slow-down for ImageNet is about $13{\times}$.
This comparison demonstrates the performance penalty of approximate multiplication simulation plus the use of a custom CUDA library.
Essentially, the difference in bold-black and bold-blue slow-downs is due to the overheads of the approximate multiplier simulation.
The slow-down numbers for System-II are slightly lower than for System-I for training and inference in general since V100 GPU has Tensor Cores, which are faster than the architecture in the GTX1080.

Previous work TFapprox with \textit{inference}-only framework~\cite{vaverka_tfapprox_2020} has shown $10{\times}$ slowdown for small dataset-architectures, despite only supporting 8-bit integer datatype. ApproxTrain supports floating-point \textit{training} and \textit{inference} with just $7.32{\times}$ slow-down ($7.32{\times}$ is the geometric mean of all experiments containing both AMDENSE and AMCONV2D operators). Despite 16-bit FP datatype (bfloat16) is used to benchmark ApproxTrain, we compare inference performance on approximate CONV2D operators of ApproxTrain and TFapprox. We reproduce the TFapprox project~\cite{Tfapprox} and benchmark both ApproxTrain and TFapprox with identical measurement procedure in System-II (GTX1080 GPU). As shown in Figure \ref{fig:run_time_perf_cmp}, similar inference performance can be observed for both ApproxTrain and TFapprox across 4 different dataset-architectures, containing intensive approximate CONV2D operations. Note that, TFapprox only supports 8-bit integer inference for approximate CONV2D operator while ApproxTrain enables generic (1, \textit{e}, \textit{m}) FP \textit{training} and \textit{inference} at once for both AMCONV2D and AMDENSE operators.


\smallskip

\subsubsection{\textbf{ApproxTrain GPU performance vs CPU-based approximate multiplier simulations}}

We compare the performance of ApproxTrain with GPU (ATxG) with the runtimes of approximate multiplier training/inference on CPU (ATxC).
The speed-ups of these comparisons are highlighted in bold (green) in Table~\ref{tab_timingtraining}\&\ref{tab_timinginference}. We observe that for training on System-I, the geometric mean speed-up is more than $2500$x! Similarly, for inference, the speed-up is more than $2869$x. The speed-ups for System-II are slightly lower as the GPU is less powerful while CPUs in System-II have similar performance to System-I.
Thus, the presented ApproxTrain offers a fast and easy solution for testing approximate multipliers and DNNs compared to naive simulations on CPU.

\section{Conclusions}
\label{sec_conc}

This paper proposed a framework (ApproxTrain) to perform training and inference with approximate FP multipliers through simulation. Firstly, a novel flow is proposed to effortlessly convert C/C++-based functional models of the approximate multipliers into optimal AMSim. Then, this AMSim is integrated into ApproxTrain (extension of Tensorflow), leveraging CUDA to speed up the simulation. ApproxTrain allows researchers to flexibly evaluate and explore their approximate multiplier designs in various DNNs. Our evaluations show that approximate multipliers (AFM) could converge DNNs as well as FP32 and bfloat16 multipliers. The GPU run-time shows significant speedup over CPU run-times, making it practically feasible. ApproxTrain is released as open-source~\cite{AMDNN} for further contributions from the research community.
\section{Acknowledgement}
This research/project was undertaken with the assistance of resources and services from the National Computational Infrastructure (NCI), which is supported by the Australian Government. We acknowledge Yikai Wang for monitoring experiments.

\balance
\bibliographystyle{plain}
\bibliography{reference} 

\end{document}